\documentclass[pra,aps,twocolumn,superscriptaddress]{revtex4}
\usepackage{}
\usepackage{lipsum}
\usepackage{graphicx}  
\usepackage{epstopdf}
\usepackage{dcolumn}   
\usepackage{bm}        
\usepackage{amssymb}   
\usepackage{amsmath}   
\usepackage{amsthm}
\usepackage{chemarrow}
\usepackage{color}
\usepackage{mathrsfs}
\usepackage{float}
\usepackage{bbold}
\usepackage{bbm}
\usepackage[a4paper,colorlinks=true,
linkcolor=blue,citecolor=blue,
pdfauthor={ },
pdftitle={ },
pdfsubject={ },
pdfkeywords={ }]{hyperref}
\theoremstyle{plain}

\newtheorem*{theorem*}{Theorem}

\newtheorem{theorem}{Theorem}

\newtheorem{itlemma}{Lemma}[section]
\newtheorem{itproposition}[itlemma]{Proposition}
\newtheorem{itcorollary}[itlemma]{Corollary}
\newtheorem{itremark}[itlemma]{Remark}
\newtheorem{itremarks}[itlemma]{Remarks}
\newtheorem{itdefinition}[itlemma]{Definition}
\newtheorem{itexample}[itlemma]{Example}

\newenvironment{lemma}{\begin{itlemma}\rm}{\end{itlemma}} 
\newenvironment{remark}{\begin{itremark}\rm}{\end{itremark}} 
\newenvironment{remarks}{\begin{itremarks} \rm}{\end{itremarks}}
\newenvironment{corollary}{\begin{itcorollary}\rm}{\end{itcorollary}}
\newenvironment{proposition}{\begin{itproposition}\rm}{\end{itproposition}}
\newenvironment{definition}{\begin{itdefinition}\rm}{\end{itdefinition}}
\newenvironment{example}{\begin{itexample}\rm}{\end{itexample}}
\newenvironment{fact}{\noindent {{\bf Fact}}:\ \ }{\hfill \medskip}
\newenvironment{claim}{\noindent {\em Claim}. \ \ }{\hfill \medskip}
\newcommand{\CC}{\mbox{${\rm \:  C\!\!\! I
			\;\;}$}}
\newcommand{\be}[1]{\begin{equation}\label{#1}}
\newcommand{\ee}{\end{equation}}
\newcommand{\bl}[1]{\begin{lemma}\label{#1}}
	\newcommand{\br}[1]{\begin{remark}\label{#1}}
		\newcommand{\brs}[1]{\begin{remarks}\label{#1}}
			\newcommand{\bt}[1]{\begin{theorem}\label{#1}}
				\newcommand{\bd}[1]{\begin{definition}\label{#1}}
					\newcommand{\bp}[1]{\begin{proposition}\label{#1}}
						\newcommand{\bc}[1]{\begin{corollary}\label{#1}}
							\newcommand{\bfact}[1]{\begin{fact}\label{#1}}
								\newcommand{\bex}[1]{\begin{example}\label{#1}}
									\newcommand{\ec}{\end{corollary}}
								\newcommand{\efact}{\end{fact}}
							\newcommand{\eex}{\end{example}}
						\newcommand{\el}{\end{lemma}}
					\newcommand{\er}{\end{remark}}
				\newcommand{\ers}{\end{remarks}}
			\newcommand{\et}{\end{theorem}}
		\newcommand{\ed}{\end{definition}}
	\newcommand{\ep}{\end{proposition}}
\newcommand{\epr}{\end{proof}}
\newcommand{\bpr}{\begin{proof}}
\newcommand{\bcl}{\begin{claim}}
	\newcommand{\ecl}{\end{claim}}

\newcommand{\bi}{\begin{itemize}}
	\newcommand{\ei}{\end{itemize}}
\newcommand{\ben}{\begin{enumerate}}
	\newcommand{\een}{\end{enumerate}}
\newcommand{\figref}[1]{\figurename~\ref{#1}}

\begin{document}


\title{Time-optimal Control of Independent Spin-$\frac{1}{2}$ Systems under {Simultaneous}  Control}

\date{\today}

\author{Yunlan Ji}
\affiliation{
CAS Key Laboratory of Microscale Magnetic Resonance and Department of Modern Physics, University of Science and Technology of China, Hefei, Anhui 230026, China}
\affiliation{
	Synergetic Innovation Center of Quantum Information and Quantum Physics, University of Science and Technology of China, Hefei, Anhui 230026, China}

\author{Ji Bian}
\affiliation{
CAS Key Laboratory of Microscale Magnetic Resonance and Department of Modern Physics, University of Science and Technology of China, Hefei, Anhui 230026, China}
\affiliation{
	Synergetic Innovation Center of Quantum Information and Quantum Physics, University of Science and Technology of China, Hefei, Anhui 230026, China}

\author{Min Jiang}
\affiliation{
CAS Key Laboratory of Microscale Magnetic Resonance and Department of Modern Physics, University of Science and Technology of China, Hefei, Anhui 230026, China}
\affiliation{
	Synergetic Innovation Center of Quantum Information and Quantum Physics, University of Science and Technology of China, Hefei, Anhui 230026, China}

\author{Domenico D'Alessandro}
\email[]{daless@iastate.edu}
\affiliation{
Department of Mathematics, Iowa State University, Ames, Iowa 50011, USA}

\author{Xinhua Peng}
\email[]{xhpeng@ustc.edu.cn}
\affiliation{
CAS Key Laboratory of Microscale Magnetic Resonance and Department of Modern Physics, University of Science and Technology of China, Hefei, Anhui 230026, China}
\affiliation{
Synergetic Innovation Center of Quantum Information and Quantum Physics, University of Science and Technology of China, Hefei, Anhui 230026, China}
\affiliation{
Synergetic Innovation Center for Quantum Effects and Applications, Hunan Normal University, Changsha 410081, China}


\begin{abstract}
We derive the explicit solution of the problem of time-optimal control by {a common} magnetic fields for two independent spin-$\frac{1}{2}$  particles.
Our  approach is  based on the Pontryagin Maximum Principle and  a novel  symmetry reduction technique.
We experimentally implement the optimal control using zero-field nuclear magnetic resonance. This reveals an average gate error of $1\%$ and a $70 \%$ to $80$ $\%$ decrease in the experiment duration as compared to existing methods.  This is the first {\it{analytical}} solution and {\it{experimental}} demonstration of time-optimal control in such a system and it provides a route to achieve time optimal control in {more general  quantum systems}.  
\end{abstract}


\maketitle
\section{Introduction}
Time-optimal control (TOC) problems in quantum systems are ubiquitous and important in multiple applications \cite{Shapiro2003,Schulte2005,Ernst1990, glasereuro}.
Because  the inevitable noise from the environment degrades quantum states and  operations over time,
inducing  quantum dynamics  in  minimal time utilizing TOC becomes a preferable choice.
Different mathematical approaches exist to obtain accurate TOC protocols \cite{Carlini2006,  Rezakhani2009, Wang2015, Geng2016}, with the Pontryagin Maximum Principle (PMP) \cite{Sachkov} unifying some of them. However, independently of the method used,  analytic solutions are rare in optimal control and often a  numerical prescription is given for the optimal control law with known problems of convergence to the actual solution \cite{Rao,Chen2015}.    Previous works mainly consider time optimization with controls which address spin individually  \cite{Khaneja2002, Yuan2005}. However this is  difficult in many experiments, and  {control of all spins  simultaneously affected} is a common scenario.

In this paper, we use the Pontryagin Maximum Principle (PMP) (see, e.g., \cite{Sachkov}) and a novel symmetry reduction technique \cite{Albertini2018},  to obtain the optimal control laws for a system of  two uncoupled spin-$\frac{1}{2}$ particles, under simultaneous  control.  Our symmetry reduction technique allows us to reduce the number of 
unknown parameters and to obtain {\it analytic} solutions. 
We implemented the TOC law using zero-field NMR  \cite{Blanchard2016}, obtaining experimental fidelity as high as $99\%$ and a gain of about $70\% \sim 80\%$ in the experiment time over previously known schemes \cite{Bian2017,Jiang2017}.

In particular, our model is as follows: Two spin-$\frac{1}{2}$ particles with different gyromagnetic ratios $\gamma_1$ and $\gamma_2$ are simultaneously subject to a global {(spatially uniform)}  control field
$\vec u(t):=u_{x,y,z}$. The Hamiltonian is
$H(\vec u)= \sum_{j=x,y,z} (\gamma_1  \sigma_{j} \otimes {\bf 1}_2 +\gamma_2 {\bf 1}_2 \otimes   \sigma_j )u_j $,
where $\vec{\sigma}=\sigma_{x,y,z}$ are the Pauli matrices,  {
\begin{equation*}
\sigma_x=  \begin{pmatrix} 0 & 1 \cr 1 & 0 \end{pmatrix}, \qquad
\sigma_y=  \begin{pmatrix} 0 & -i \cr i & 0 \end{pmatrix}, \qquad
\sigma_z=  \begin{pmatrix} 1 & 0 \cr 0 & -1 \end{pmatrix}, \qquad
\end{equation*}}
\noindent and ${\bf 1}_n$ denotes  the $n \times n$ identity.  The  
problem is to steer the identity ${\bf 1}_4 \in \textrm{SU}(4)$ to any desired matrix
$U_{f,1} \otimes U_{f,2} \in \textrm{SU}(2) \otimes \textrm{SU}(2)$ under a constraint of the form $|\vec u|  \leq D $, $D\not=0$. 
We assume $\gamma_1\not=\gamma_2$ (heteronuclear spins) \cite{f1}  which implies {\it controllability}  \cite{Mikobook} in $\textrm{SU}(2) \otimes \textrm{SU}(2)$, { that is, every operator in $\textrm{SU}(2) \otimes \textrm{SU}(2)$ can be reached with an appropriate (arbitrarily bounded) control.} 
Let ${\cal B}$ be the subspace in $\textrm{su}(2)\oplus \textrm{su}(2)$ defined as
$
{\cal B}:=\texttt{span} \{ i\sigma_x \otimes {\bf 1}_2+\gamma {\bf 1}_2 \otimes i\sigma_x,
-i\sigma_y \otimes {\bf 1}_2-\gamma {\bf 1}_2 \otimes i\sigma_y, i\sigma_z \otimes {\bf 1}_2+\gamma {\bf 1}_2 \otimes i\sigma_z \}
$ with $\gamma := \gamma_2/\gamma_1$.
The problem of optimal control can be stated  as finding a function
$X:=X(t)=-iH(\vec u)$ with values in ${\cal B}$, where $H$ is the above Hamiltonian, 
so that the solution of the {\it Schr\"odinger operator equation}, 
\begin{equation}\label{SCROP}
\dot U=X(t)U \qquad U(0)={\bf 1}_4, 
\end{equation}
 reaches $U_{f,1} \otimes U_{f,2}$  in minimum
time. Using the Hamiltonian of the system, we have  
$X=-i\sum_{j=x,y,z} (\gamma_1  \sigma_{j} \otimes {\bf 1}_2 +\gamma_2 {\bf 1}_2 \otimes   \sigma_j )u_j$ and  the knowledge of $X=X(t)$ is equivalent to the knowledge of the control $\vec u$.  Moreover the bound on the control in the optimal control problem implies a bound on the norm of $X$. In particular $|\vec u| \leq D$ if and only if 
\be{BoundX}
\| X(t) \| \leq \tilde{L}:= |\gamma_1| \sqrt{1+\gamma^2}D
\ee 
for all $t$. [The inner product in $\textrm{su}(n)$ (in particular for $n=4$) is  $\langle A, B \rangle:=\frac{1}{n}\textrm{Tr}(AB^\dagger)$ so that $\| A \|:=\frac{1}{\sqrt{n}}\sqrt{\text{Tr}(AA^\dagger)}$.]

This paper is organized as follows: 
In section \ref{sec_3} we describe the method to obtain TOC, i.e., PMP and symmetry reduction technique. A step-by-step protocol, and a flow chart (\figref{flow_1}) illustrating the algorithm to obtain TOC are also summarized in this section.
In section $\textrm{\ref{sec_4}}$, in preparation to the experiments we carried out, we consider the special case where we want to perform a rotation on the first spin while leaving the second spin unchanged. We prove that, in this case, the core step of the proposed algorithm amounts to an integer optimization problem with constraints (Theorem \ref{OptimizationProblem}).  We apply this method to our experiment in section \ref{sec_5}, together with an evaluation of the quality of the control. The conclusion and discussion of this work is given in Section \ref{sec_6}. Some useful computations and extra considerations are collected in the appendix.

\section{Time-optimal control law} \label{sec_3}
We   combine the PMP on Lie groups \cite{Sachkov} with the use of 
symmetry reduction \cite{Albertini2018}. This results in an algorithm to obtain the optimal control laws.

\subsection{The PMP and the form of optimal control}\label{SUBS1} The following theorem which uses the Pontryagin Maximum Principle for systems on Lie groups \cite{Sachkov} gives a description of the functional form of the optimal control and trajectory.  
\bt{Theoptcontra}
Write  $X=-i\sum_{j=x,y,z} (\sigma_{j} \otimes {\bf 1}_2 +\gamma {\bf 1}_2 \otimes   \sigma_j )u_j:=X_1 \otimes {\bf 1}_2 + \gamma {\bf 1}_2 \otimes X_1$ (so that $X_1:=\sum_{j=x,y,z} -i \sigma_j u_j$) the optimal control, and  $U:=U_1\otimes U_2$ the optimal trajectory. Then there exist matrices $P$ and $A$ in $\textrm{su}(2)$ such that
\begin{equation}\label{U15}
X_1=  e^{At} P {e^{-At}}, \: 
U_1= e^{At} e^{(P-A)t}, \:   U_2=e^{At} e^{(\gamma P-A)t}.
\end{equation}
\et
{\it Proof.} Consider the controlled dynamics (\ref{SCROP}). Conditions given by PMP for right invariant systems on Lie groups \cite{Sachkov} say that if $X=X(t)$ and $U=U(t)$ is an optimal pair of control  
and trajectory, respectively, in time $t_{\textrm{min}}$,  then the following facts hold true: There exists a nonzero pair $(F,\lambda_0)$ with $F$ in the associated Lie 
algebra, in this case $\textrm{su}(2) \oplus \textrm{su}(2)$,   and $\lambda_0$ a scalar such that, defined the (PMP) Hamiltonian $\tilde H (F,\hat U,V):=\langle F,\hat U^\dagger V \hat U\rangle$, with $\hat U$ in the given Lie group, in this case  
$ \textrm{SU}(2) \otimes \textrm{SU}(2)$, and $V$ in the control set, in this case $V \in {\cal B}$, so that, for almost every $t \in [0,t_{\textrm{min}}]$,
\be{io}
\tilde H (F,U(t),X(t))=\max_{\|V\| \leq \tilde{L}} \tilde H (F,U(t),V)=\lambda_0.
\ee
By applying the Goh condition (see, e.g., Appendix C in \cite{OptRes} and references therein) it also follows that $\lambda_0\not=0$ and therefore $F \not=0$ as well \cite{footnote1}.

Define $F:=F_1\otimes {\mathbf{1}_2} +{\mathbf{1}_2} \otimes F_2 \in \textrm{su}(2) \oplus \textrm{su}(2) $, and recall $U:=U_1 \otimes U_2$ and $X:=X_1 \otimes {\bf 1}_2+ \gamma {\bf 1}_2 \otimes X_1$. 
We have
\begin{small}
	\begin{equation}
	\label{conto}
	\begin{aligned}
	&\tilde H(F,U,X):=\langle F, U^\dagger X U\rangle \\
	&=\langle   U_1 F_1 U_1^\dagger \otimes {\bf 1}_2 + {\bf 1}_2 \otimes U_2 F_2 U_2^\dagger, X_1  \otimes {\bf 1}_2 + \gamma
	{\bf 1}_2 \otimes  X_1 \rangle\\
	&= \langle
	\tilde B + \gamma \tilde C, X_1  \rangle,
	\end{aligned}
	\end{equation}
\end{small}
where we used
\be{TBTC}
\tilde B:=U_1 F_1 U_1^{\dagger},
\qquad \tilde C:=U_2 F_2 U_2^\dagger.
\ee
Furthermore, 
$\tilde B+\gamma \tilde C$ is
never zero since this would imply $\lambda_0=0$ in the PMP. From the 
Cauchy-Schwartz inequality for
the inner product $\langle A, B\rangle=\frac{1}{2}\textrm{Tr}(AB^\dagger)$ in $\textrm{su}(2)$ and the bound on $X$ which gives a constant bound on $X_1$, from (\ref{io})  there exists a constant $k$ such that
$X_1:=\frac{k (\tilde B+\gamma \tilde C)}{\| \tilde B + \gamma \tilde C \|}$ (recall that the norm of $X$ and therefore the norm of $X_1$ is constant \cite{footnote2}). Replacing this 
into the last one of (\ref{conto}), we have
\be{quasiul}
\tilde{H}(F,U,X)=2k\| \tilde B+\gamma \tilde C \|=\lambda_0 \not=0,
\ee
which implies that $\| \tilde B+\gamma \tilde C \|$ is constant. Therefore denoting by $B$ and $C$ the matrix functions obtained from  $\tilde B$ and $\tilde C$  in (\ref{TBTC}) by
possibly  re-scaling  $F_1$ and $F_2$, we have, for the form of the optimal control,
\be{U1}
X=X_1 \otimes {\mathbf{1}}+ \gamma {\mathbf{1}} \otimes X_1, \quad \texttt{with} \quad
X_1=B+\gamma C, 
\ee
and 
\begin{equation}\label{F1set}
	B:=B(t)=U_1(t) \hat F_1 U_1^\dagger(t), 
\end{equation}
\begin{equation} \label{F2set}
	C:=C(t)=U_2(t) \hat F_2 U_2^\dagger(t), 
\end{equation}
for matrices $\hat F_1$ and $\hat F_2$ (rescaled $F_1$ and $F_2$) in $\textrm{su}(2)$.
The above derived optimal control
candidates in (\ref{U1}) are  in `feedback form', that is,
they  depend on the current value of the state of the system, $U_1 \otimes U_2$. We now transform them  into the explicit form given in the statement of the theorem. 
From \eqref{SCROP},  using (\ref{U1}), we have that the optimal 
trajectory is $U_1\otimes U_2$ with 
\be{X1}
\dot U_1=X_1 U_1, \qquad U_1(0)={\mathbf{1}}_2,
\ee
\be{X2}
\dot U_2=\gamma X_1 U_2, \qquad U_2(0)={\mathbf{1}}_2.
\ee
Using (\ref{X1})  and differentiating $B$ in (\ref{F1set}), we obtain with (\ref{X1}) 
\be{BBB}
\dot B=[X_1, B],  \qquad B(0)=\hat F_1,
\ee
and from $X_1$ in (\ref{U1}), we have
\be{BBB2}
\dot B=\gamma [C,B], \qquad B(0)=\hat F_1,
\ee
Analogously for $C$ we obtain
\be{CCC}
\dot C=\gamma [B,C],  \qquad C(0)=\hat F_2.
\ee
By combining (\ref{BBB2}) and (\ref{CCC}), we have that
\be{dBdC}
\dot B+ \dot C\equiv 0.
\ee
Therefore $B+C=\tilde A$, for a constant $\tilde A \in \textrm{su}(2)$. Therefore, from (\ref{U1}) we have
\be{U11}
X_1(t)=(1-\gamma) B(t) + \gamma \tilde A.
\ee
Replacing this in (\ref{BBB}) and solving we obtain that $B(t)=e^{\gamma \tilde A t} \hat F_1 e^{-\gamma \tilde A t}$, which replaced in (\ref{U11}) gives
\be{U111}
X_1(t)=e^{\gamma \tilde  A t} \left((1-\gamma) \hat F_1+ \gamma \tilde A \right)
e^{- \gamma \tilde  A t}.
\ee
By choosing $A:=\gamma \tilde A$ and $P=(1-\gamma)\hat M_1+ \gamma \tilde A $, and solving \eqref{X1}, \eqref{X2}, one obtains:
\begin{eqnarray}
&X_1=  e^{At} P {e^{-At}},\,
U_1= e^{At} e^{(P-A)t},\, \nonumber \\
&U_2=e^{At} e^{(\gamma P-A)t},\nonumber
\end{eqnarray}
which are formula (\ref{U15}). This completes the proof of the theorem.

Theorem \ref{Theoptcontra} reduces the search of the optimal control on a space of functions to the search for the matrices $A$ and $P$ in $\textrm{su}(2)$. 
Using Theorem \ref{Theoptcontra}, the TOC problem is then to 
find two matrices $A$ and $P$ in $\textrm{su}(2)$ and the minimum of $t$,  such that 
$U(t)= e^{At} e^{(P-A)t} \otimes e^{At} e^{(\gamma P-A)t}
=U_{f,1} \otimes U_{f,2}$, 
for desired final conditions $U_{f,1}$ and $U_{f,2}$ for systems $1$ and $2$, respectively. From the theorem we need $U_1(t)=e^{At} e^{(P-A)t}=\pm U_{f,1}$ and $U_2(t)=e^{At}e^{(\gamma P-A)t}=\pm U_{f,2}$, with minimum $t$. The only constraint on $A$ and $P$ in $\textrm{su}(2)$  is (cf. (\ref{BoundX}))
\begin{equation}\label{pnorm}
\begin{aligned}
\|P\|&=\|X_1\|=\frac{1}{\sqrt{1+\gamma^2}}\|X\|\\
&:=\tilde L / \sqrt{1+\gamma^2}=|\gamma_1|D:=L,
\end{aligned}
\end{equation}
which is a consequence of the bound on the control. This results in {\it six} parameters to be chosen. However, a reduction of 
the number of parameters can be achieved using the symmetry of the problem \cite{Albertini2018} as explained next. 

\br{limgamma}
The matrices $A$ and $P$ in the theorem, which are used in the expression of the optimal control, depend on the parameter $\gamma$. This is true also for the minimum time $t$.  At the limit  $\gamma \rightarrow 1$ the system (\ref{SCROP}) loses controllability on $\textrm{SU}(2)\otimes \textrm{SU}(2)$ in that the operations at the limit are the same on the two spins. This implies that, except for very special final conditions,  the minimum time goes to $\infty$ as $\gamma$ tends to $1$. This can be seen from the expressions of the trajectories $U_1$ and $U_2$ in (\ref{U15}). We can write $U_1U_2^\dagger$ as 
$$
U_1U_2^\dagger=e^{At}e^{(P-A)t} e^{(-\gamma P+A)t} e^{-At}=
$$
$$
e^{At}e^{(P-A)t}e^{(-P+A)t+(1-\gamma)Pt} e^{-At}. 
$$ 
This has to be a fixed matrix for a desired final condition. If, by contradiction, we assume that $t=t(\gamma)$ is bounded as $\gamma$ goes to $1$ the term $(1-\gamma)Pt$ will tend to zero (since $P$ is also bounded because of the bound on the control). Therefore the matrix on the right hand side will tend to the identity which is true only for final operations equal to each other on the two systems. 
\er

\subsection{Symmetry reduction}\label{SUBS2} Let $G$ be the Lie subgroup of $\textrm{SU}(2) \otimes \textrm{SU}(2)$ of matrices of the form $Y\otimes Y$ with $Y \in \textrm{SU}(2)$. The Lie group  $G$ acts on $\textrm{SU}(2) \otimes \textrm{SU}(2)$ by conjugation, i.e., for $U \in \textrm{SU}(2) \otimes \textrm{SU}(2)$, $Y\otimes Y \in G$,  as 
$U \rightarrow (Y \otimes Y) U (Y^\dagger  \otimes Y^\dagger)$. With this action,  
$G$ is a group of {\it symmetries} for the time optimal control  problem in the sense of \cite{Albertini2018}. This implies that if $X:=X(t)$ is an optimal control 
and $U:=U(t)$ is the corresponding  optimal trajectory for a final condition $U_f$, then,  for any fixed  
$Y \otimes Y \in G$, $(Y \otimes Y)  X (Y^\dagger  \otimes Y^\dagger)$ is an optimal 
 control and  $(Y \otimes Y) U (Y^\dagger \otimes Y^\dagger)$ the corresponding optimal trajectory for the final condition 
$(Y \otimes Y) U_f( Y^\dagger \otimes Y^\dagger)$ with the same minimum time. A direct way to see this is to consider equation (\ref{SCROP}) with $X$ the optimal control to reach $U_f$, defining $\hat Y:=Y \otimes Y$. By multiplying (\ref{SCROP}) on the left by $\hat Y$ and on the right by $\hat Y^\dagger$ we obtain a differential equation for $\hat Y U \hat Y^\dagger$. $\hat Y X \hat Y^\dagger$ is again an admissible control since it still belongs to ${\cal B}$ and its norm is the same as the one of $X$. Therefore we have an admissible control which drives to $\hat Y U_f \hat Y^\dagger$ in the same time as the minimum time to drive to $U_f$. Moreover this is the minimum time also for $\hat Y U_f \hat Y^\dagger$. If there was a shorter time, this would imply (repeating the argument) a shorter time for $U_f$ which contradicts optimality. Summarizing, if we find the optimal control and trajectory for a final condition $U_f$, we have found the optimal control and trajectory for any final condition of the type $(Y \otimes Y) U_f (Y^\dagger \otimes Y^\dagger)$. $U_f$ and $(Y \otimes Y) U_f (Y^\dagger \otimes Y^\dagger)$ are said to be in the same {\it orbit}. It is sufficient to find the optimal control and trajectory for one  representative in the desired  orbit in order to find optimal controls for all elements in the orbit. We can use this fact to reduce the number of parameters we look for in the optimal control.

Given $A$ and $P$ in $\textrm{su}(2)$,  choose $Y_1 \in \textrm{SU}(2)$ diagonalizing  $A$ ($Y_1A{Y_1}^{\dagger}=i \omega \sigma_z$). Also let $Y_2\in \textrm{SU}(2)$ diagonal and such that
$Y_2 Y_1 P Y_1^{\dagger} Y_2^\dagger= i(a \sigma_z-b\sigma_y)$
with $\sqrt{a^2+b^2}=L$ (cf. (\ref{pnorm})).
Then $U_{1} \otimes U_{2}$ from Theorem \ref{Theoptcontra} is in the same orbit as $\tilde U := \tilde U_1 \otimes \tilde U_2:=YU_1Y^\dagger \otimes YU_2Y^\dagger$ with $Y:=Y_1 Y_2$. Here:
\begin{equation}
\begin{aligned}
&Y U_1 Y^\dagger  = Y e^{At} e^{(P -A)t}Y^\dagger= e^{i\omega \sigma_z t} e^{(ia \sigma_z -ib \sigma_y -i\omega \sigma_z)t}, \\ 
&Y U_2 Y^\dagger   = Y e^{At} e^{(\gamma P -A)t} Y^\dagger  = e^{i\omega \sigma_z t} e^{(i\gamma a \sigma_z - i\gamma b \sigma_y -i\omega \sigma_z)t}.
\label{U12}
\end{aligned}
\end{equation}
Consequently the (TOC)  problem of searching for {\it six} parameters is reduced to the problem of searching for {\it three} independent parameters, i.e., $\omega$, $t_{min}$ and $(a,b)$ with $(\sqrt{a^2+b^2}=L)$. The parameter $t_{\textrm{min}}$ is the minimum time $t$ such that $YU_1 Y^\dagger \otimes YU_2 Y^\dagger$, in (\ref{U12})  is in the same orbit as the desired $U_{f,1} \otimes U_{f,2}$. 

The control $(Y\otimes Y) X (Y^\dagger \otimes Y^\dagger)=YX_1 Y^\dagger\otimes {\bf 1}_2+\gamma {\bf 1}_2 \otimes YX_1 Y^\dagger$, with   $Y X_1 Y^\dagger=e^{i\omega \sigma_z t}
(ia \sigma_z-ib\sigma_y) e^{-i\omega \sigma_z t}$ drives the state optimally from the identity ${\bf 1}_4$ to an element in the same orbit as
$ U_{f,1} \otimes  U_{f,2}$. 
The problem is therefore split in two. First one chooses   the parameters $\omega$, $a$ ($b$) and $t=t_{\textrm{min}}$ to reach the orbit of $U_{f,1} \otimes U_{f,2}$ and then one `adjusts' via a similarity transformation of the form  $Y\otimes Y$ to obtain {\it exactly} the desired final condition $U_{f,1} \otimes U_{f,2}$.

In order to follow this procedure we need an explicit  description of the space of orbits, $SU(2) \otimes SU(2)/G$, the {\it orbit space} \cite{footnote3}. To simplify the problem,  we slightly relax the equivalence relation on $\textrm{SU}(2)\otimes \textrm{SU}(2)$ to a relation, $\sim_\times$, on $\textrm{SU}(2)\times \textrm{SU}(2)$, so that  $(W_1,Z_1) \sim_\times  (W_2,Z_2)$ if and only if  there exists a $Y$ in $\textrm{SU}(2)$ such that  
$(W_2,Z_2)=(YW_1 Y^\dagger, YZ_1Y^\dagger)$. We have that $W_1 \otimes Z_1$ 
and $W_2 \otimes Z_2$ are in the same orbit if and only if 
$(W_1,Z_1)\sim_\times (W_2,Z_2)$ {\it or}  $(W_1,Z_1)\sim_\times (-W_2,-Z_2)$, i.e., two points in the orbit space $\textrm{SU}(2)\times \textrm{SU}(2)/\sim_\times$ correspond  to the same point in the orbit space $SU(2) \otimes SU(2)/G$. The characterization of $\textrm{SU}(2)\otimes \textrm{SU}(2)/\sim_\times$ is given in the following proposition. The proof is reported in the Appendix section \ref{proofchar}  \cite{voronov}. Here ${\bf D}$ denotes the closed unit disc in the complex plane, that is, the set of complex numbers $x \in \CC$ such that $|x| \leq 1$.

\begin{figure}[t]
	\centering
	\includegraphics[width=0.4\textwidth, height=0.2\textheight]{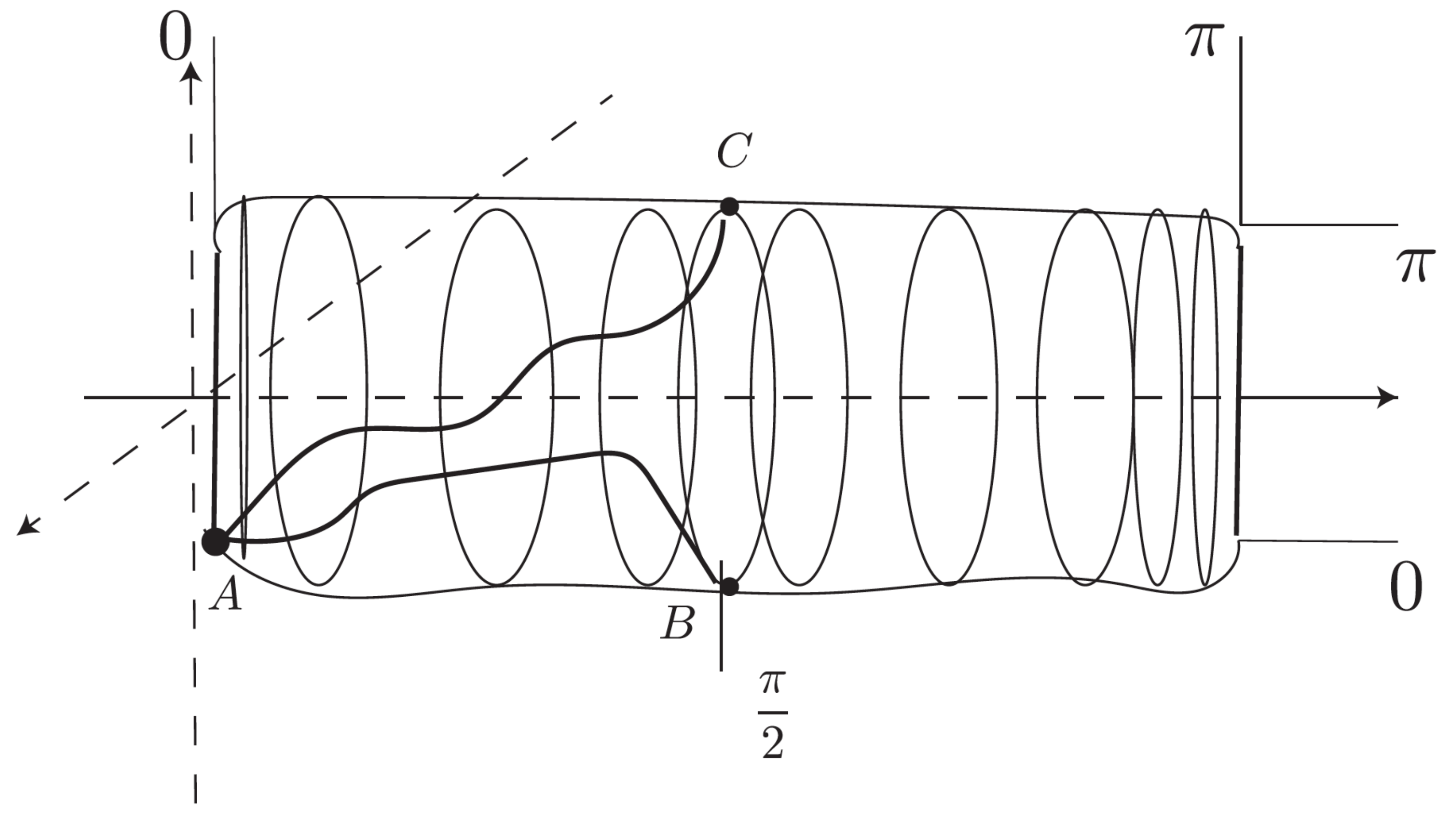}
	\caption{Representation of the Orbit Space $\textrm{SU}(2) \times \textrm{SU}(2)/\sim_\times$. The point $A$ corresponds to the equivalence class $[({\bf 1}_2, {\bf 1}_2)]$; The point $B$ corresponds to the equivalence class $[(\textrm{Phase}, {\bf 1_2})]$ with $\textrm{Phase}$ gate defined in (\ref{DefiNOT1}) and $C$ corresponds to the equivalence class $[(\textrm{Phase}, -{\bf 1_2})]$. Trajectories are depicted in the orbit space joining $A$ and $B$, $A$ and $C$.}
	\label{Figura1}
\end{figure}

\bp{SU2orbits}
There exists a one to one and onto map $\Psi$
\be{Psimap}
\begin{aligned}
	\Psi  : &(\textrm{SU}(2)\times \textrm{SU}(2))/\sim_\times\longrightarrow \\
	&\left((0,\pi) \times {\bf D} \right) \bigcup \left( \{0 \} \times [0,\pi] \right)
	\bigcup \left( \{ \pi\} \times [0,\pi] \right),
\end{aligned}
\ee
defined as follows:
\begin{enumerate}
	\item If $U$ has an eigenvalue $e^{i\phi}$, with $\phi \in (0,\pi)$ and therefore $U:=S\Lambda S^\dagger$ with $S \in \textrm{SU}(2)$ and $\Lambda:=\begin{pmatrix} e^{i\phi} & 0 \cr 0 & e^{-i\phi} \end{pmatrix}$ then
	\be{Psi1}
	\Psi\left([(U,Z)]\right)=(\phi, (S^\dagger Z S)_{1,1}) \in (0,\pi) \times {\bf D},
	\ee
	where $(S^\dagger Z S)_{1,1}$ denotes the $(1,1)$ entry of the matrix $S^\dagger Z S$ which is an element of ${\bf D}$, and $[(U,Z)]$ denotes the orbit with representative $(U,Z)$ ($[(U,Z)] \in (\textrm{SU}(2) \times \textrm{SU}(2))/\sim_\times$).
	
	\item If $U$ is the identity matrix ${\bf 1}_2$ and $e^{i\psi}$ with $\psi \in [0,\pi]$ is an eigenvalue of $Z$ then
	\be{Psi2}
	\Psi\left([(U,Z)]\right)=(0,\psi) \in \{0 \} \times [0,\pi].
	\ee
	
	\item  If $U$ is the negative of the identity matrix, i.e., $-{\bf 1}_2$,  and  $e^{i\psi}$ with $\psi \in [0,\pi]$ is an eigenvalue of $Z$ then
	\be{Psi3}
	\Psi\left([(U,Z)]\right)=(\pi,\psi) \in \{\pi \} \times [0,\pi].
	\ee
\end{enumerate}
\ep
Topologically therefore  the orbit space $\textrm{SU}(2) \times \textrm{SU}(2)/\sim_\times$ looks like a deformed  {\it solid cylinder} as in \figref{Figura1}, where the  discs at the left and right ends are  degenerate to a segment ($[0,\pi]$), and every other cross section is (homeomorphic to) a disc. 

Since $\Psi$ in the proposition is a bijection  $(U_1, Z_1) \sim_\times (U_2, Z_2)$ if and only if $\Psi([(U_1, Z_1)])=\Psi([(U_2, Z_2)])$. Therefore a test for $\sim_\times$ equivalence is given in the following corollary. 
\bc{Orbits} $(U_1,Z_1)\sim_\times (U_2,Z_2)$ if and only if one of the 
following occurs: 1) $U_1=U_2=\pm {\bf 1}_2$ and the spectrum of $Z_1$ is equal 
to the spectrum of $Z_2$; 2) $U_1:=S_1 \Lambda S_1^\dagger$ and 
$ U_2:=S_2 \Lambda S_2^\dagger$ for the same diagonal $\Lambda$ \cite{theorm2note}, (with $S_1$ and $S_2$ in $\textrm{SU}(2)$) and $(S_1^\dagger Z_1 S_1)_{1,1}=(S_2^\dagger Z_2 S_2)_{1,1}$ where $R_{1,1}$ denotes the $(1,1)$ entry of the matrix $R$. 
\ec 
{\it Proof.} The case 1 of the corollary corresponds to the cases 2 and 3 of the proposition. The case 2 of the corollary corresponds to the case 1 of the proposition.  

Analysis on quotient spaces in the context of quantum optimal control 
was also done in \cite{Khan1}. The quotient space in \cite{Khan1} is a {\it symmetric space} \cite{Helgason} while the one described in the above proposition and corollary is a {\it stratified space} \cite{Snyaticki}.

\subsection{Procedure to obtain the time optimal  control}\label{Proce}

\begin{widetext}

\begin{figure*}[b]  
	\includegraphics[scale=0.5]{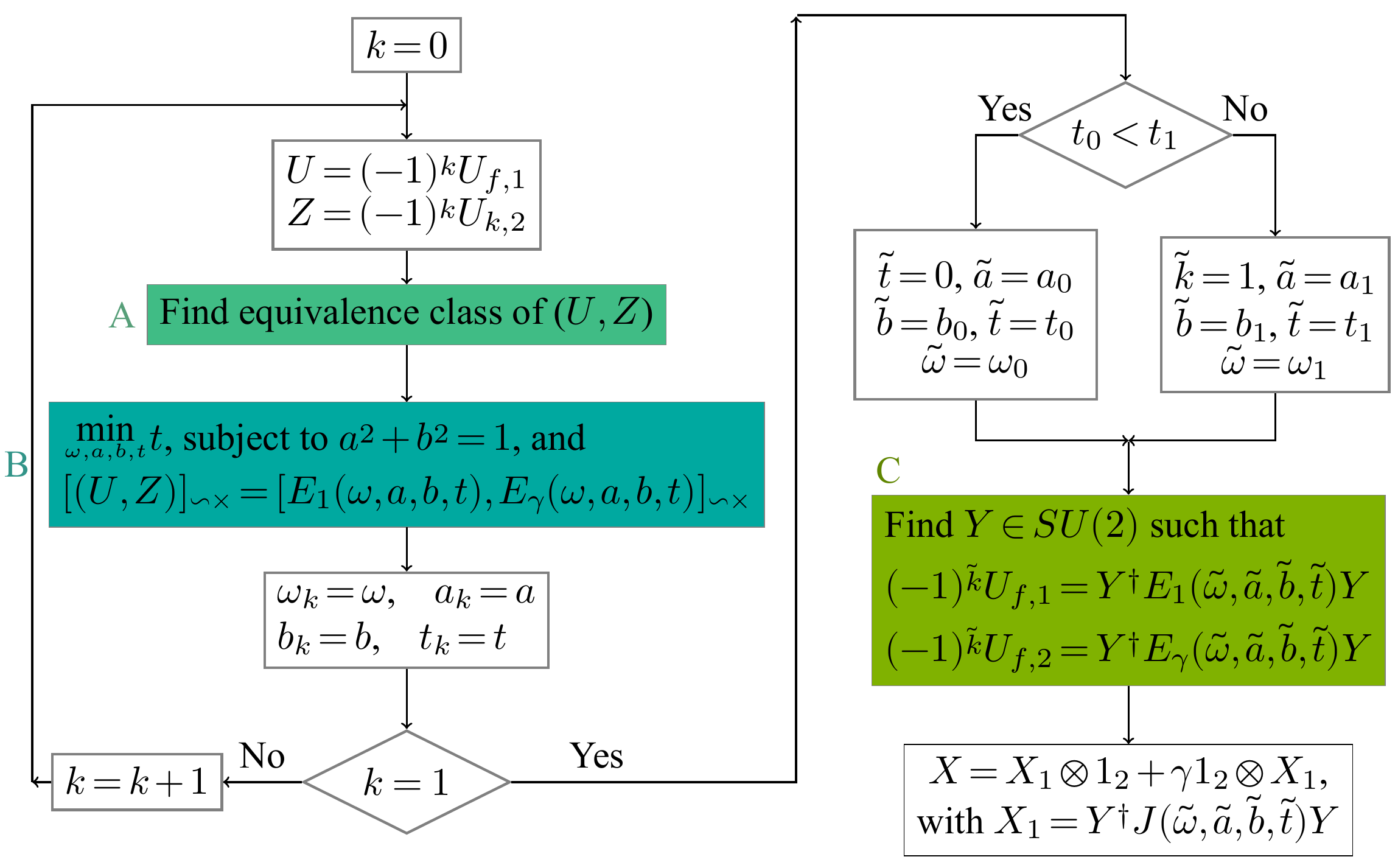}
\caption{Flow chart of the procedure to find the optimal control $X=X_1 \otimes {\bf 1}_2+ \gamma {\bf 1} \otimes X_1$. For brevity, we use the notation $E_\gamma(\omega, a, b,t):=e^{i\omega \sigma_z t} e^{i(\gamma a \sigma_z-\gamma b \sigma_y-\omega \sigma_z)t}$ and $J(\omega, a, b, t):=e^{i\omega \sigma_z t}(ia \sigma_z - ib \sigma_y)e^{-i \omega \sigma_z t}$.   }
	\label{flow_1}
\end{figure*}

\end{widetext}

Combining the explicit form of the optimal control obtained in subsection \ref{SUBS1} with the symmetry reduction described in subsection \ref{SUBS2}, the  main points of the {\it protocol} to find the optimal control field can be summarized as follows.  
%
\begin{enumerate}	
	 \item The PMP gives the form of optimal control $X=\sum_{j=x,y,z}-i(\sigma_{j} \otimes {\bf 1}+ \gamma {\bf 1} \otimes \sigma_{j})u_j=X_1\otimes {\bf 1}+ \gamma {\bf 1} \otimes X_1$, and trajectory, $U(t)=U_1(t) \otimes U_2(t)$. These are given by $X_1=e^{At}Pe^{-At}$, $U_1=e^{At}e^{(P-A)t}$,   $U_2=e^{At}e^{(\gamma P-A)t}$, where  $A$ and $P$ are constants in $\textrm{su}(2)$, parametrized therefore by $6$ real parameters.
	 
	 \item Symmetry reduction further reduces the number of unknown parameters to 3: If the control $X$ and trajectory $U$ are an optimal pair with optimal time $t_{\textrm{min}}$, then $\hat YX\hat Y^{\dagger}$ and $\hat YU\hat Y^{\dagger}$ is also an optimal pair with $t_{\textrm{min}}$ (\ref{U12}). There are, $\omega, t_{\textrm{min}}, a$ $(b)$, with $\sqrt{a^2+b^2}=L$, 3 unknown parameters to be determined now ($\hat Y$ is also unknown, but it can be determined if the above three parameters  are known).
	
	\item Find real values  $\omega$, $t=t_{\textrm{min}}$, $a$ ($b$)
	(see point 1 above  and (\ref{U12})), such that 
	$$\qquad (e^{i\omega \sigma_z t} e^{(ia \sigma_z -ib \sigma_y -i\omega \sigma_z)t},e^{i\omega \sigma_z t} e^{(i\gamma a \sigma_z - 
		i\gamma b \sigma_y -i\omega \sigma_z)t})$$
		 is in the same class as $(U_{f,1},U_{f,2})$ (Use Corollary \ref{Orbits}), and $t$ is minimum. 
	
	 \item Repeat step 3 with the substitution: $(U_{f,1},U_{f,2})\rightarrow(-U_{f,1},-U_{f,2})$. Choose the minimum time between these two cases, and the corresponding $\omega, t:=t_{\textrm{min}}, a$ $(b)$.
	
	\item Find $Y \in \textrm{SU}(2)$ such that 
	\begin{equation*}
	\begin{aligned}
	\pm U_{f,1}=Y^{\dagger}e^{i\omega\sigma_zt}e^{(ia\sigma_z-ib\sigma_y-i\omega\sigma_z)t}Y,\\
	\pm U_{f,2}=Y^{\dagger}e^{i\omega\sigma_zt}e^{(i\gamma a\sigma_z-i\gamma b\sigma_y-i\omega\sigma_z)t}Y,
	\end{aligned}
\end{equation*}	
with $t=t_{\textrm{min}}$. The $\pm$  sign is chosen according to step 4.	 
	 
	 \item The optimal control is $X:=X_1 \otimes {\bf 1}_2+\gamma {\bf 1}_2 \otimes X_1$ with 
	$X_1:=Y^\dagger e^{i \omega \sigma_z t}(ia\sigma_z-ib \sigma_y)  
	e^{-i \omega \sigma_z t} Y$.

\end{enumerate} 
 A simplification in the above procedure is obtained assuming for $L$ in (\ref{pnorm}) 
$L:=\sqrt{a^2+ b^2}=1$, that is, normalizing the bound $D$ on the control. 
We can always recover the optimal control for the original problem. In fact,  
if $\vec u_N(t)$ is the optimal control with  a bound $\frac{1}{|\gamma_1|}$ ($L=1$) in time $t_N$,  
$\vec u(t)=L \vec u_N(Lt)=
|\gamma_1|D u_N(|\gamma_1| D t)$ will be an optimal control for a general  bound $D$, in minimum time $\frac{t_N}{L}$. 
We shall therefore set $L=1$ in the following discussion. 
FIG. \ref{flow_1} gives a flow chart of the algorithm to find the optimal control. This algorithm takes  the desired final condition $U_{f,1} \otimes U_{f,2}$ as the input and obtains the time optimal control. 

\vspace{0.25cm}

The routine  $A$ of the flow chart is carried out using corollary \ref{Orbits} and proposition \ref{SU2orbits}. It amounts to a standard eigenvalue-eigenvector problem for which in general 
there are many available numerical algorithms and in our case can be solved by hand since we are dealing with $2 \times 2$ matrices. Task $C$ is the solution of a system of linear equations in which we can use any available parametrization of matrices  in $SU(2)$. It corresponds to step $5$ of the previously described protocol. The solution of the minimization problem in $B$ which corresponds to task $3$ of the protocol is arguably the most difficult step of the algorithm and the core of our solution method. We shall often refer in the following to this step as `Task 3'. One method to tackle this step  is   to use the concept of {\it reachable set} as discussed in \cite{Albertini2018}. Consider the geometric description of the orbit space 
$\textrm{SU}(2) \times \textrm{SU}(2)/\sim_\times$ given in \figref{Figura1}. This can be depicted (from Corollary \ref{Orbits}) as a {\it solid cylinder} where the two extreme discs are collapsed to a segment.  The orbit of a pair $(\pm U_{f,1}, \pm U_{f,2})$ is a point in this space. If we fix $t$ and vary $\omega$ and $a$ ($b$),  we obtain a surface in this $3-$D space which is the boundary of the set of orbits reachable at time $t$. The first $t$ such that this surface includes the desired point is the minimum time. The values of $\omega$ and $a$ ($b$) where the intersection occurs are the optimal values for the parameters. 
Alternatively one can  directly tackle the optimization problem to minimize $t$ under the constraint that the pair 
\be{pair2}
(e^{i\omega \sigma_z t} e^{i(a \sigma_z-b \sigma_y-\omega \sigma_z)t}, e^{i\omega \sigma_z t} e^{i(\gamma a \sigma_z-\gamma b \sigma_y-\omega \sigma_z)t})
\ee
 is in the same equivalence class (with respect to the equivalence relation $\sim_{\times}$) as $ (U_{f,1}, U_{f,2})$, using the explicit computations of matrix exponentials which are reported in the Appendix (section \ref{UsC}).   This problem is in some cases simplified and can be solved analytically, as in the application  to our experiments which we discuss next.

\section{Application to selective single-spin rotations} \label{sec_4}
We applied the above procedure to obtain 
the TOC law for  selective rotations 
on the first  spin, i.e., $U_f = U_{f,1} \otimes U_{f,2} = e^{-i \mathbf{n}\cdot \vec{\sigma} \frac{\theta}{2}} \otimes \mathbf{1}_2$, where  $\mathbf{n}$ is a unit vector and the rotation angle $\theta$ is chosen in  $(0,2 \pi)$. The possible final 
 conditions in $\textrm{SU}(2) \times \textrm{SU}(2)$ are $(e^{-i \mathbf{n}\cdot \vec{\sigma} \frac{\theta}{2}} , {\bf 1}_2)$ and $(-e^{-i \mathbf{n}\cdot \vec{\sigma} \frac{\theta}{2}} , -{\bf 1}_2)$. 
Since
$U_{f,2}=\pm \mathbf{1}_2$ is invariant under similarity transformations, the parameters in step 3 of the above procedure ( routine $B$ in FIG. \ref{flow_1})  have to be chosen so that in (\ref{U12}) $e^{i\omega \sigma_z t} e^{(i\gamma a \sigma_z - i\gamma b \sigma_y -i\omega \sigma_z)t}=\pm \mathbf{1}_2$. This implies 
that $i \gamma a \sigma_z-i \gamma b \sigma_y-i\omega \sigma_z t$, and therefore $b \sigma_y$,  commutes with $e^{-i \omega \sigma_z t}$ which is diagonal. Therefore  
$e^{-i \omega \sigma_z t}=\pm {\bf 1}_2$ or $b=0$.
If $b=0$ ($a=\pm1$), the final conditions give $\gamma t=k\pi$ and $\cos(t)=(-1)^k \cos(\frac{\theta}{2})$.  So, in this case, we choose as candidate minimum time $t=|\frac{k}{\gamma}|\pi$ where $|k|$ is the smallest integer (if any) such that $\cos( | \frac{k}{\gamma}|\pi)=(-1)^k \cos(\frac{\theta}{2})$. If $b \not=0$ ($|a|<1$), the problem to minimize $t$ subject to the final condition (i.e., Task 3) can be  transformed into an integer optimization problem  as summarized in the following theorem: 
\bt{OptimizationProblem}
Assume that the optimal parameters are such that $b \not=0$ in Task 3. Define the function 
\begin{equation}\label{emmegamma}
M_\gamma(s,m,l,k):=m^2(1-\gamma)+(s\frac{\theta}{2 \pi}+l)^2\gamma-k^2,  
\end{equation}
where $s=\pm 1$ and $m,l,k$ are integers with $m,k>0$, and $l,k$ share the same parity if $\theta\not=\pi$. When $s=1$, $l \geq 0$; when $s=-1$, $l > 0$.
Then the minimum time $t$, $t_{min}$,  is the minimum value of 
$\pi \sqrt{\frac{M_\gamma (s,m,l,k)}{\gamma(1-\gamma)}}$,
with the constraint 
\be{constr55}
[m-s\frac{\theta}{2\pi}-l]^2 < \frac{M_\gamma (s,m,l,k)}{\gamma (1-\gamma)} < [m+s\frac{\theta}{2 \pi}+l]^2 . 
\ee
With, $t=t_{min}$ and  the optimal $(s,m,l,k)$ the corresponding parameters $(\omega, a, b)$ are given by $\omega=\frac{m\pi}{t}$, $a=\frac{\omega}{2\gamma}+\frac{\gamma}{2\omega} - \frac{k^2 \pi}{2 t m\gamma}$, 
and $b=\pm \sqrt{1-a^2}$
\et 
{\it Proof.} Using the calculation of exponential of matrices in section \ref{UsC},  
and the conditions on the final state for the first spin, we obtain
\be{cond1}
(e^{i\omega t}+e^{-i\omega t}) \cos(\eta_1 t) = \pm 2 \cos{(\frac{\theta}{2})}. 
\ee
This is in particular obtained taking the trace of the matrix in (\ref{X1iii}) and imposing that it is equal to ($\pm$) the trace of the desired final  condition. 
By imposing that the matrix in (\ref{X2iii2}) is equal to $\pm$ the identity, we get 
\be{cond2}
e^{i\omega t} (\cos(\eta_\gamma t)+
i\frac{\gamma a-\omega }{\eta_\gamma} \sin(\eta_\gamma  t))
=\pm 1. 
\ee
In these formulas, we used the definitions 
$\eta_1:=\sqrt{\omega^2+1-2a\omega}$,
$\eta_\gamma:=\sqrt{\omega^2+\gamma^2-2a\omega \gamma}$.
From \eqref{cond2}, since $|a|<1$ and recall the definition of
$\eta_\gamma$,
we obtain the two conditions
\be{Req1}
\omega t=m\pi,
\ee
\be{Req2}
\eta_\gamma t:=\sqrt{\omega^2 +\gamma^2 -2\gamma a \omega} \,t=k\pi,
\ee
for integers $m$ and $k>0$.
Using (\ref{Req1}) in (\ref{cond1}), and the fact that whether we use $+$ or $-$ depends on $(-1)^m(-1)^k$ from (\ref{Req2}), (\ref{Req1}) and (\ref{cond2})  we obtain 
$
(-1)^m 2\cos(\eta_1 t)=2(-1)^m (-1)^k\cos(\frac{\theta}{2}) 
$, that is, 
\begin{equation}\label{ops}
\cos(\eta_1 t)=(-1)^k\cos(\frac{\theta}{2}).
\end{equation}
Therefore, in particular we have 
\begin{equation}\label{Req3}
 \eta_1t:=\sqrt{\omega^2+1 -2a\omega} \,t=s\frac{\theta}{2}+l\pi, \\
\end{equation}
with $l$ and $k$ having the same parity if $\theta\not=\pi$. Therefore the condition on the final state is verified if and only if conditions (\ref{Req1}), (\ref{Req2}) and (\ref{Req3}) are verified with $l$ and $k$ having the same parity if $\theta\not=\pi$.

If $\omega=0$ then $m=0$ from (\ref{Req1}), and (\ref{Req2}), (\ref{Req3}) give $t =\frac{|k|}{|\gamma|} \pi $ and $\cos(t)=(-1)^k\cos(\frac{\theta}{2})$.  So the situation is the same as the one discussed for $b=0$ and therefore we can avoid considering this case as for now we assume that the optimal occurs (only) for $b \not=0$.
Therefore we assume $\omega \not=0$.   We can in fact assume $\omega > 0$, and therefore $m>0$ also, since for every pair $(\omega, a)$ satisfying (\ref{Req1}), (\ref{Req2}), (\ref{Req3}) for a certain $t$, the pair $(-\omega, -a)$ also satisfies (\ref{Req1}), (\ref{Req2}), (\ref{Req3}) with the same $t$.

Now using (\ref{Req1}) in (\ref{Req2}) and (\ref{Req3}), we obtain 
\be{Req5}
m \sqrt{\omega^2 +\gamma^2-2a\gamma \omega}=\omega k,
\ee
and 
\begin{equation}\label{Req31}
m\sqrt{\omega^2+1 -2a\omega} = s \frac{q}{2} \omega +l \omega.
\end{equation}
with $\theta=q\pi, q=(0,2)$. We have that if $s=1$, $l \geq 0$, while if $s=-1$, $l$ must be $>0$. Eliminating  $a\omega$ by combining equations  (\ref{Req31}) and (\ref{Req5}), we obtain that
\be{omega}
\omega^2=\frac{m^2 \gamma (1-\gamma)}{M_\gamma (s,m,l,k)},
\ee
with $M_\gamma(s,m,l,k) \gamma(1-\gamma) > 0.$ ($M_\gamma(s,m,l,k)$ is defined in \ref{emmegamma})

For a certain quadruple  $(s,m,l,k)$, the time is then (from (\ref{Req1}))
\be{time}
t=\frac{m\pi}{\omega}=\pi \sqrt{\frac{M_\gamma}{\gamma(1-\gamma)}}.
\ee
Using $\omega=\frac{m\pi}{t}$ in (\ref{Req3}),
we obtain
\be{Req7}
\pi^2\left[ m^2 -(s\frac{q}{2}+l)^2 \right] +t^2=2 a m \pi t,
\ee
The condition $|a|<1$ is equivalent  to the fact that the absolute
value of the right hand side of \eqref{Req7} is strictly less than $2m\pi t$.
By setting
\be{Req8}
\left\{\pi^2\left[ m^2 -(s\frac{q}{2}+l)^2 \right] +t^2\right\}^2 < 4m^2 \pi ^2 t^2,
\ee
we obtain the condition for $t^2$,
\be{Req9}
[m\pi-(s\frac{q}{2}+l)\pi]^2 < t^2 < [m\pi+(s\frac{q}{2}+l)\pi]^2.
\ee
Replacing (\ref{time}) in this, we have
\be{Req10}
[m-s\frac{q}{2}-l]^2 < \frac{M_\gamma(s,m,l,k)}{\gamma (1-\gamma)} < [m+s\frac{q}{2}+l]^2,
\ee
which is the same as  (\ref{constr55}) if we recall that $\theta=q\pi$.   We remark that this is in fact the only condition since the left inequality in (\ref{Req10}) implies $M_\gamma(s,m,l,k) \gamma(1-\gamma) > 0$. Using the values of $\omega$ and $t$ in (\ref{Req2}) one obtains the expression for $a$ (and therefore $b$) in the statement of the theorem.  
Theorem \ref{OptimizationProblem} transforms Task 3 of the procedure into 
an integer optimization problem. 
\br{Rema0}
Given the particular final condition, the sign of $b$ in the statement of the theorem is arbitrary. It does not affect the eigenvalues of the transformation on the first spin given that the transformation on the second spin is the identity.  
\er

\br{Rema1}
From the proof of the theorem it follows that we have simultaneously considered the case $U\otimes {\bf 1}$ and $(-U) \otimes (-{\bf 1})$ therefore we do not need, in this case, to perform  the test `$t_0 < t_1$' in the algorithm of FIG.\ref{flow_1}, and we can directly move on 
to the task in $C$ of the flow chart, in this case.     
\er
There is no general algorithm  to solve the optimization problem of Theorem \ref{OptimizationProblem} which treats $\gamma$ as a free parameter. However when $\gamma$ is given a numerical value, such a problem can usually be solved. One possible 
technique is to use a  $min-min$ strategy as follows: First for given $l$ and $m$ one 
finds the minimum or maximum (according to the sign of $\gamma(1-\gamma)$)  value of $k$ (with the same parity of $l$) so that condition (\ref{Req10}) is verified. This is because the minimization of $t$ corresponds to the minimization of $\frac{M_\gamma}{\gamma(1-\gamma)}$ (from (\ref{time}))  and $M_\gamma$ is given in (\ref{emmegamma}). Such an optimal  $k$  will be a function of $l$ and $m$. Then  one finds $l$ and $m$ to minimize $t$ in (\ref{time}). Such a procedure might have to be repeated for $s=1$ and $s=-1$ and the optimal times compared. As an illustration of this technique we consider the case $\gamma=\frac{1}{2}$ in the appendix (section \ref{gamma12}). 
Alternatively, one can apply enumeration or numerical search in the space of
$(m,l,k)$ to get an optimal candidate, then independently prove its time optimality. This is the technique we have used in our experimental implementation as it is described in the next section.

\section{Experiments in zero-field NMR} \label{sec_5}

At zero field, all spins have identical (zero) Larmor frequencies and they cannot be addressed by separate control fields.
They can be manipulated by applying pulsed magnetic fields $ \vec B $ along three directions acting on all the spins. The main advantage of zero-field NMR is that it does not need superconducting magnets. This makes the experiment set-up more flexible compared to high-field NMR. When the control fields satisfy the condition $\gamma_{1,2}| \vec B| \gg  |2\pi J|$ (the general case in liquid-state NMR experiments),
where $J$ is the spin-spin coupling constant, the spin systems at zero field behave as independent spins in simultaneous  control.

\begin{figure}[t]
	\centering
	\includegraphics[width=0.4\textwidth, height=0.25\textheight]{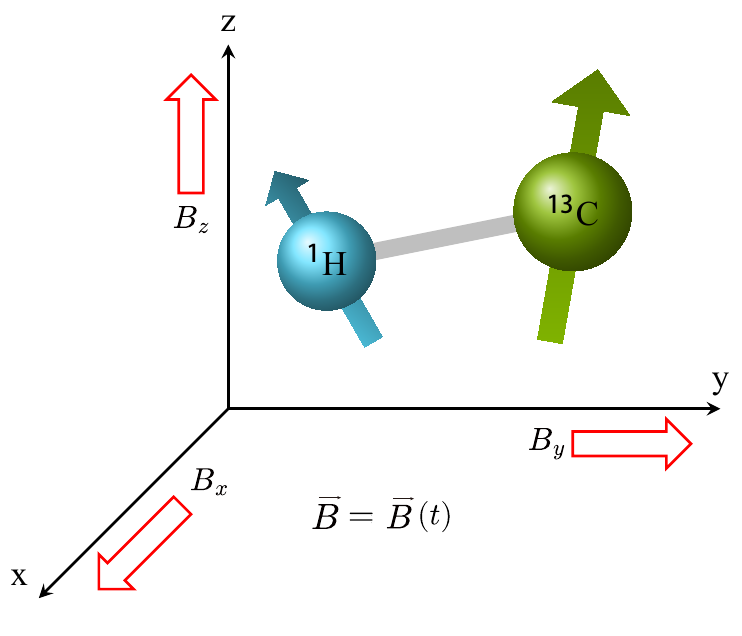}
	\caption{Schematic representation of the control via a spatially uniform magnetic field 
	$\vec B$ of the $^1\textrm{H}$-$^{13}\textrm{C}$ system.}
	\label{addcart}
\end{figure}

We experimentally demonstrated the above TOC pulses for an $^1$H-$^{13}$C system,
i.e., $^{13}$C-formic acid ($^1$H-$^{13}$COOH), at zero field. This system is schematically depicted in FIG. \ref{addcart}. 
The $^1$H-$^{13}$C spin-spin constant is $J\approx 221.9~\textrm{Hz}$ and the lifetime of singlet-triplet coherence \cite{Emondts2014} is $T_2\approx2.0~\textrm{s}$. For the $^1$H-$^{13}$C system,
$ \frac{\gamma_\textrm{C}}{\gamma_\textrm{H}} \approx 0.2514=\gamma$. We describe next how to solve the optimization problem of Theorem \ref{OptimizationProblem} and obtain the TOC for this value of $\gamma$.

\subsection{Determination of TOC}

We now determine the time optimal control for the case of $\gamma:=0.2514$ which corresponds to our experiment, and therefore this value of $\gamma$ will be assumed in this subsection.
We start with an ansatz  for  $M_\gamma$ (and therefore $t_{min}$) 
solving the optimization problem of Theorem \ref{OptimizationProblem}. This is given by  $s=l=k=m=1$, i.e., $M_\gamma(1,1,1,1)$. It  is achieved by enumeration in a small range [in the space of $(s,m,l,k)$]. However its  time optimality  cannot be proved by simple enumeration since the triple in $(m,l,k)$ is  not in a 
bounded range. 
\subsubsection{Proof of optimality of $M_{\gamma}(1,1,1,1)$}\label{proofopt}

We only consider  the case with $q \in (0,1]$ (recall the definition $\theta=q \pi$) since the case $q \in (1,2)$ can be treated similarly. We show that there is no admissible quadruple $(s,m,l,k)$ which gives a value of $M_{\gamma}(s,m,l,k)$ strictly less than $M_{\gamma}(1,1,1,1)$. There are two subcases to consider $s=1$ and $s=-1$. 

{{\bf Case $s=1$}} 

 Define $\delta:= m-l$ in $M_\gamma(s,m,l,k)=M_\gamma(1,m,l,k)$. We first observe the following fact: 

\bl{Lemmad}
If $M_\gamma(1,m,l,k)<M_\gamma(1,1,1,1)$ then $\delta=0$ or $\delta=1$
\el

{\it Proof.} Using $M_\gamma(1,m,l,k)<M_\gamma(1,1,1,1)$ the lower bound $(m-l-\frac{q}{2})^2$ in \eqref{Req10} gives
\begin{equation}
\label{s1con1}
(m-l-\frac{q}{2})^2<\frac{M_\gamma(1,1,1,1)}{\gamma (1-\gamma)}.
\end{equation}
Direct computation of $M_\gamma(1,1,1,1)$ gives $M_\gamma(1,1,1,1)=\left(\frac{q^2}{4} +q \right)\gamma$, using this in (\ref{s1con1}) and setting $\delta:=m-l$, we get 
\begin{equation}
\label{s1con2}
-\frac{\sqrt{\frac{q^2}{4}+q}}{\sqrt{1-\gamma}}+\frac{q}{2}<\delta<\frac{\sqrt{\frac{q^2}{4}+q}}{\sqrt{1-\gamma}}+\frac{q}{2}. 
\end{equation}
This leads to a restriction on the possible values of $\delta$ (recall $\gamma=0.2514$) as claimed:
\begin{eqnarray}\label{secondcase}
\delta=
\left \{ \begin{array}{ll}
0 \qquad &0  < q \leq \frac{2(-8743+50\sqrt{32458})}{1257} (\approx 0.42)\\

0\ \text{or} \ 1 \quad &\frac{2(-8743+50\sqrt{32458})}{1257}  < q \leq 1
\end{array} \right.
\end{eqnarray}

The following two propositions consider the two cases $\delta=0$ and $\delta=1$ and show that, in these cases,  there is no quadruple $(1,m,l,k)$ such that $M_\gamma(1,m,l,k)< M_\gamma(1,1,1,1)$, so completing the proof for the case $s=1$. 

\bp{d0A} Assume $\delta:=m-l=0$. Then $M_\gamma(1,m,l,k)=M_\gamma(1,m,m,k) \geq 
M_\gamma(1,1,1,1)$, for any admissible values of $m$, $l$ and $k$. 
\ep 

{\it Proof.} If $\delta=0$, $m=l$ gives $M_\gamma=m^2-k^2+ \frac{q^2}{4} \gamma+mq\gamma$. 
Since $M_\gamma>0$ we must have $m \geq k$. In fact, 
if $m<k$, we would have 
$$M_\gamma=m^2+qm\gamma-k^2+\frac{q^2}{4}\gamma \leq (k-1)^2+q(k-1)\gamma-k^2+\frac{q^2}{4}\gamma \leq 
$$
$$-3+q\gamma+\frac{q^2}{4}\gamma<0,$$ 
which is a contradiction.  Set $\epsilon:=m-k$  
$(0 \leq \epsilon<m)$, $\epsilon$ integer. Assuming by contradiction that 
$M_\gamma(1,m,l,k)< M_\gamma(1,1,1,1)$ with the additional requirement that $m=l$ leads to the inequality 
$$
m^2-k^2+mq\gamma < q \gamma. 
$$
Replacing $k$ with $m-\epsilon$, after some algebraic manipulations, we obtain 
\begin{equation}
\epsilon<m<\frac{q\gamma+\epsilon^2}{q\gamma+2\epsilon},
\end{equation}
which leads to:
\begin{equation}
\label{s1d00}
\epsilon^2+q\gamma\epsilon-q\gamma<0.
\end{equation}
From \eqref{s1d00}, we have that the only possibility is $\epsilon=0$. Therefore $m=k$.  Together with $\delta=0$, this indicates  $m=l=k$. However 
$$M_\gamma (1,m,m,m)=\frac{q^2}{4}\gamma+qm\gamma \geq \frac{q^2}{4}\gamma+q\gamma=M_\gamma(1,1,1,1),$$
with equality valid if and only if $m=1$.  Therefore  no smaller $M_\gamma$ can be  found in this case.

\bp{D1B} Assume $\delta:=m-l=1$. Then $M_\gamma(1,m,l,k)=M_\gamma(1,m,m-1,k) \geq 
M_\gamma(1,1,1,1)$, for any admissible values of $m$, $l$ and $k$. 
\ep 

{\it Proof.} If $\delta=1$ (corresponding to the second case in (\ref{secondcase})), by using $M_\gamma$ in (\ref{emmegamma}) with $l=m-1$,   $M_\gamma>0$ becomes:
\begin{equation}\label{s1d1}
m^2-k^2+m(q\gamma-2\gamma)+\gamma+\frac{q^2}{4}\gamma-q\gamma>0
\end{equation}
Since
\begin{small}
\begin{equation}
\begin{aligned} m(q\gamma-2\gamma)+\gamma+\frac{q^2}{4}\gamma-q\gamma&<q\gamma-2\gamma+\gamma+\frac{q^2}{4}\gamma-q\gamma\\
&=-\gamma+\frac{q^2}{4}\gamma<0,
\end{aligned}
\end{equation}
\end{small}
we must have $m>k$ to make \eqref{s1d1} hold. Set $k=m-\epsilon$  ($0<\epsilon<m$, $\epsilon$  integer). 
Then assuming by contradiction $M_\gamma(1,m,l,k)<M_\gamma(1,1,1,1)$ (cf. (\ref{emmegamma})), with $l=m-\delta=m-1$, we obtain 
\begin{equation}
m(2\epsilon+q\gamma-2\gamma)<2q\gamma-\gamma+\epsilon^2.
\end{equation}
Thus:
\begin{equation}
\epsilon<m<\frac{2q\gamma-\gamma+\epsilon^2}{2\epsilon+q\gamma-2\gamma}.
\end{equation}
From the condition $\frac{2(-8743+50\sqrt{32458})}{1257}  < q \leq 1$, we have:
\begin{equation}
\begin{aligned}
-1&<\frac{2\gamma-q\gamma-\sqrt{(q\gamma-2\gamma)^2-4(\gamma-2q\gamma)}}{2}<\epsilon<\\
&\frac{2\gamma-q\gamma+\sqrt{(q\gamma-2\gamma)^2-4(\gamma-2q\gamma)}}{2}<1
\end{aligned}
\end{equation}
So no $\epsilon$ satisfies this requirement. 

We have thus shown 
for the case $s=1$ that $M_\gamma(1,1,1,1)$ is the minimum.

{\bf Case $s=-1$}

The following lemma analogous to Lemma  \ref{Lemmad} says that there are two cases again to consider. We set again $\delta:=m-l$.

\bl{Lemmadplus}
If $M_\gamma(-1,m,l,k)<M_\gamma(1,1,1,1)$ then $\delta=0$ or $\delta=-1$
\el

{\it Proof.} From the constraint (\ref{Req10}) on $M_\gamma$ written  for $s=-1$, we know that if $M_\gamma(s,m,l,k)$ is strictly less than $M_\gamma(1,1,1,1)$, with 
the lower bound (\ref{Req10}) now equal to  $(m-l+\frac{q}{2})^2$, we must have:
\begin{equation}
\label{sf1con1}
(m-l+\frac{q}{2})^2<\frac{M_\gamma (s,m,l,k)}{\gamma (1-\gamma)} < \frac{M_\gamma(1,1,1,1)}{\gamma (1-\gamma)}.
\end{equation}
Inequality (\ref{sf1con1}) gives:
\begin{equation}
\label{sf1con2}
-\frac{\sqrt{\frac{q^2}{4}+q}}{\sqrt{1-\gamma}}-\frac{q}{2}<\delta<\frac{\sqrt{\frac{q^2}{4}+q}}{\sqrt{1-\gamma}}-\frac{q}{2},
\end{equation}
which leads to the restrictions on $\delta$ (recall $\gamma=0.2514$):
\begin{eqnarray}\label{tobor}
\delta=
\left \{ \begin{array}{ll}
0 \qquad &0  < q \leq \frac{2(-8743+50\sqrt{32458})}{1257} \\

-1 \ \text{or} \ 0 \qquad &\frac{2(-8743+50\sqrt{32458})}{1257}  < q \leq 1
\end{array} \right.
\end{eqnarray}

The following two propositions consider the cases $\delta=0$ and $\delta=-1$ separately and show that it is not possible in these cases that $M_\gamma(-1,m,l,k)< M_\gamma(1,1,1,1)$. This is analogous to what has been done in Propositions \ref{d0A} and \ref{D1B} and completes all the subcases, thus showing the optimality of $M_{\gamma}(1,1,1,1)$. 
\bp{d0B} Assume $\delta:=m-l=0$. Then $M_\gamma(-1,m,l,k)=M_\gamma(-1,m,m,k) \geq 
M_\gamma(1,1,1,1)$, for any admissible values of $m$, $l$ and $k$. 
\ep

{\it Proof.} If $\delta=0$, we have 
$M_\gamma=M_\gamma(-1,m,m,k)=m^2-k^2+\frac{q^2}{4} \gamma-qm\gamma$.
From $M_\gamma>0$ we must have $m > k$. Set $\epsilon:= m-k$ ($0 < \epsilon<m)$, $\epsilon$
  integer. The assumption, by contradiction, $M_\gamma(-1,m,m,k)< M_\gamma(1,1,1,1)$ gives 
\begin{equation}
\epsilon<m<\frac{q\gamma+\epsilon^2}{2\epsilon-q\gamma},
\end{equation}
which leads to 
\begin{equation}
\label{sf1d0}
\epsilon^2-q\gamma\epsilon-q\gamma<0.
\end{equation}
From \eqref{sf1d0}, the bound on $\epsilon$ becomes:
\begin{small}
\begin{equation}
-1<\frac{q\gamma-\sqrt{q^2\gamma^2+4q\gamma}}{2}<\epsilon<\frac{q\gamma+\sqrt{q^2\gamma^2+4q\gamma}}{2}<1,
\end{equation}
\end{small}
which cannot be satisfied by any $\epsilon$ (integer $>0$).  Therefore there is 
no smaller $M_\gamma$ in this case.

\bp{DM1B} Assume $\delta:=m-l=-1$. Then $M_\gamma(-1,m,l,k)=M_\gamma(-1,m,m+1,k) \geq 
M_\gamma(1,1,1,1)$, for any admissible values of $m$, $l$ and $k$. 
\ep 

{\it Proof.} 
Set $m-\epsilon=k$ 
($\epsilon<m,$ $\epsilon $ integer). The assumption, by contradiction,  
$M_\gamma(-1,m,l,k)=M_\gamma(-1,m,m+1,k)< M_\gamma(1,1,1,1)$ gives 
\begin{equation}
m(2\epsilon+2\gamma-q\gamma)<\epsilon^2-\gamma+2q\gamma
\end{equation}
When $0 \leq \epsilon <m$, the requirement on $\epsilon$ becomes:
\begin{equation}
\epsilon<m<\frac{\epsilon^2-\gamma+2q\gamma}{2\epsilon+2\gamma-q\gamma},
\end{equation}
which can be converted to:
\begin{equation}
\epsilon^2+\epsilon(2\gamma-q\gamma)+\gamma-2q\gamma<0.
\end{equation}
But $\epsilon^2+\epsilon(2\gamma-q\gamma)+\gamma-2q\gamma>1+2\gamma-q\gamma+\gamma-2q\gamma=1+3\gamma-3q\gamma \geq 1$.
So no $\epsilon$ can be found in this case.

When $\epsilon<0$, the requirement $M(-1,m,l,k)>0$ becomes:
\begin{equation}
\label{sf1epf}
-\epsilon^2+2m\epsilon+2m\gamma-mq\gamma+\gamma+\frac{q^2}{4}\gamma-q\gamma>0.
\end{equation}
But:
\begin{equation}
\begin{aligned}
&-\epsilon^2+2m\epsilon+2m\gamma-mq\gamma+\gamma+\frac{q^2}{4}\gamma-q\gamma\\
&\leq -1-2m+2m\gamma-mq\gamma+\gamma+\frac{q^2}{4}\gamma-q\gamma\\
&\leq -1+2\gamma-2-q\gamma+\gamma+\frac{q^2}{4}\gamma-q\gamma<0,
\end{aligned}
\end{equation}
which contradicts \eqref{sf1epf}. So no value of $\epsilon$ can be found 
in this case either.


\vspace{0.25cm}

{\bf Conclusion of the proof}

\vspace{0.25cm}

 The 
 value of the minimum time is (with $\theta=q \pi$) 
\be{tmini}
t_{min}=\pi \sqrt{\frac{M_\gamma(1,1,1,1)}{\gamma(1-\gamma)}}=
\pi \sqrt{\frac{\frac{q^2}{4}+q}{1-\gamma}}. 
\ee
For the value of $\gamma=0.2514$ we are considering this 
is indeed the optimal. The parameter $b$ has to be different from zero. In fact the 
time discussed before the statement of the Theorem \ref{OptimizationProblem}, when $b=0$ 
is (when possible) $t_{b=0}:=\frac{|k|\pi}{\gamma}$, and we have from 
(\ref{tmini}) (since $q \in (0,2)$)
$$
t_{min}:=\pi \sqrt{\frac{\frac{q^2}{4}+q}{1-\gamma}} < 
\pi \sqrt{\frac{3}{1-\gamma}}< \pi \frac{1}{\gamma} \leq t_{b=0}, 
$$
which is true since for $\gamma=0.2514$, $\sqrt{\frac{3}{1-\gamma}} < \frac{1}{\gamma}$.

We remark that the proof of optimality of $M_\gamma(1,1,1,1)$ 
 holds for a {\it range} of values of $\gamma$ which {\it includes} $0.2514$.

\begin{figure}[b]  
	\makeatletter
	\def\@captype{figure}
	\makeatother
	\includegraphics[scale=0.86]{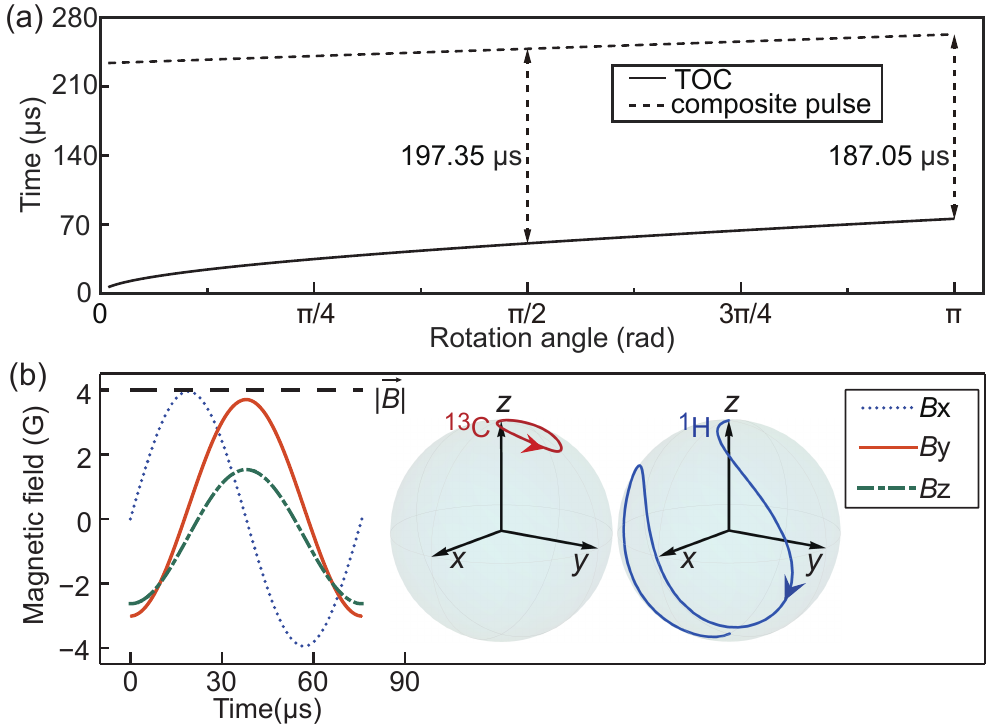}
	\caption{(color online). (a)  Comparison of time costs for 
		single-spin rotation $R^{H}_{\mathbf{n}}(\theta):= e^{-i \mathbf{n} \cdot \vec{\sigma} \frac{\theta}{2}} \otimes \mathbf{1}_2$ 
		on $^1$H in the $^1$H-$^{13}$C system between TOC and the composite-pulse sequence \cite{Bian2017, Jiang2017} implemented via
		$R^{H}_{\mathbf{n}}(\theta)=R^{C}_{\mathbf{n_{\bot}}}(\pi)e^{-iH(\vec u_0)\tau}R^{C}_{\mathbf{n_{\bot}}}(-\pi)e^{-iH(\vec u_0)\tau}$.
		Here $\vec u_0$ is a constant  field along $\mathbf{n}$ and $\tau=\frac{\theta}{2\gamma_1 |\vec u_0|}$.
		(b) Time-optimal fields and corresponding 
		trajectories on the Bloch Sphere for realizing $R_y^\textrm{H} (\pi):=e^{-i\sigma_y \frac{\pi}{2}}$.}
	\label{fig01}
\end{figure}

\begin{figure*}[t]
	\makeatletter
	
	\def\@captype{figure}
	
	\makeatother
	\centering\includegraphics[scale=0.9]{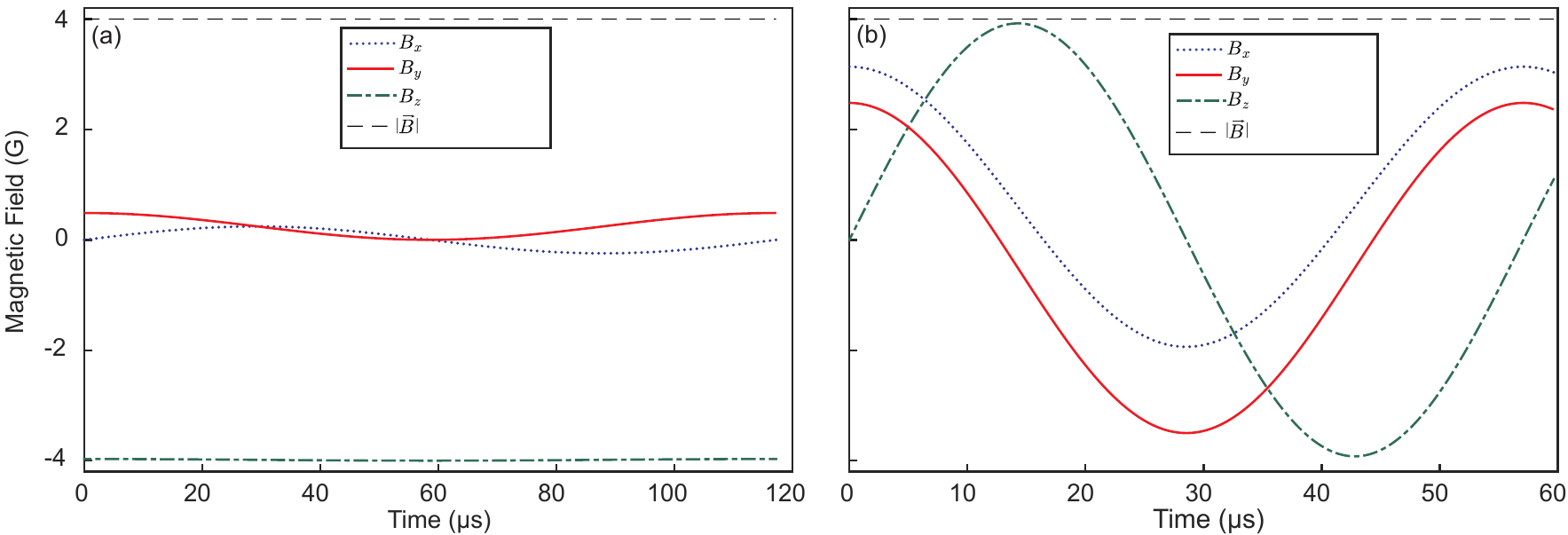}
	\caption{ (color online). TOC pulses in different $\gamma$ cases. (a) TOC $\pi$ pulse on $^{13}$C along the $y$ axis in $^1$H-$^{13}$C.  (b) TOC $\pi/2$ pulse on $^1$H along the $y$ axis in $^1$H-$^{31}$P. }
	\label{pioncinch}
\end{figure*}

\subsubsection{Explicit expression of the optimal controls}

We take as an example $\theta=\pi$. Using Theorem \ref{OptimizationProblem} we calculate the parameters ($t_{min},\omega,a,b)$ of the optimal control. We have from (\ref{emmegamma}) 
\be{emmegammaspecif}
M_\gamma=M_\gamma(1,1,1,1)=\frac{5}{4}\gamma. 
\ee  
Moreover $t_{min}={\pi} \sqrt{\frac{M_\gamma}{\gamma(1-\gamma)}}=\frac{\pi}{2} \sqrt{\frac{5}{1-\gamma}}$. We have $\omega=\frac{\pi}{t_{min}}=2 \frac{ \sqrt{1-\gamma}}{\sqrt{5}}$. We have 
(from the Theorem) 
$$
a=\frac{\omega}{2 \gamma}+\frac{\gamma}{2 \omega}- \frac{k^2 \pi}{2tm\gamma}=
\frac{\gamma}{4}\sqrt{\frac{5}{1-\gamma}}, 
$$
with $b=\pm \sqrt{1-a^2}$. With these values of $\omega, a$ and $b$, we can calculate $e^{i \omega \sigma_z t}(ia\sigma_z-ib\sigma_y) e^{-i \omega \sigma_z t}$, which gives final condition (on spin 1) $\tilde U_{f,1}=e^{i\omega \sigma_z t} e^{(i a \sigma_z -ib \sigma_y-i\omega \sigma_z)t}$ with $t=t_{min}$ (cf. (\ref{U12})). In order to complete 
Task 6 of the procedure in subsection \ref{Proce} we need to find   $Y \in \textrm{SU}(2)$ such that $U_{f,1}=Y^\dagger \tilde U_{f,1} Y$, so that $X_1=Y^\dagger e^{i \omega \sigma_z t}(ia\sigma_z-ib\sigma_y) e^{-i \omega \sigma_z t} Y$, the optimal control. The optimal control fields ($u_x,u_y,u_z$) are  obtained from $X_1=-i \sigma_x u_x-i\sigma_y u_y -i \sigma_z u_z$. If we want to consider a general bound $D$ on the control norm, we need to re-scale the optimal control which was obtained with a normalized bound 
($L:=|\gamma_1|D=1$). The re-scaling is  $\vec u(t) \rightarrow 
L \vec u(Lt)=|\gamma_1|D u(|\gamma_1| D t)$. The explicit expression  of the matrix $Y$ and the optimal control fields  are given below. These are the  control fields used in the experiment in FIG. \ref{fig01}.

\begin{small} 
\begin{equation}
{Y}=\begin{pmatrix} \frac{5}{4}\sqrt{\frac{2743}{11229}}i-\frac{1}{4}\sqrt{\frac{21257}{11229}}i & \frac{5}{4}\sqrt{\frac{2743}{11229}}+\frac{1}{4}\sqrt{\frac{21257}{11229}} \cr
-\frac{5}{4}\sqrt{\frac{2743}{11229}}-\frac{1}{4}\sqrt{\frac{21257}{11229}} & -\frac{5}{4}\sqrt{\frac{2743}{11229}}i+\frac{1}{4}\sqrt{\frac{21257}{11229}}i.
\end{pmatrix}.
\end{equation}
\end{small}
(we used in this calculations ratios of integer number to express   $\gamma=0.2514=\frac{2514}{10000}$.)
The control magnetic field $\vec{B}=(B_x,B_y,B_z):=-2(u_x,u_y,u_z) \,(|\vec{B}|=2D)$ are given by 
	\begin{equation}
	\begin{aligned}
	B_x&
	\approx
	1.98D\sin{(1.54 D \gamma_1 t)};\\
	B_y&
	\approx 0.18D-1.68D\cos{(1.54 D \gamma_1 t)};\\
	B_z&
	\approx -0.28D-1.04D\cos{(1.54 D \gamma_1 t)}.
	\end{aligned}
	\end{equation}

To show the generality of the method, we have also obtained controls for different values of $\gamma$. In particular, we have considered  $\frac{\gamma_\textrm{H}}{\gamma_\textrm{C}} \approx 3.9777=\gamma$, that is,  the rotations are implemented on $^{13}$C spin in $^1$H-$^{13}$C system (now spin 1 is $^{13}$C and spin 2 is $^{1}$H). In this case, $M_\gamma<0$ in Theorem \ref{OptimizationProblem} and it is $-M_\gamma$ that has to be minimized. We proved (like in subsection \ref{proofopt}) that  the combination $(s=-1,m=1,l=1,k=1)$ minimizes $-M_\gamma=M_\gamma(s,m,l,k)$, and the corresponding TOC $\pi$ pulse on $^{13}$C is illustrated in \figref{pioncinch} (a). We also considered $\gamma=0.4048$. This corresponds to a single-spin rotation on $^1$H in a $^1$H-$^{31}$P system. For example, a TOC $\pi/2$ pulse on $^1$H in the $^1$H-$^{31}$P system is illustrated in \figref{pioncinch} (b). In this case it is proved that the  minimal $M_\gamma=M_\gamma(s,m,l,k)$ equals $M(1,1,1,1)$. 

\begin{figure} 
	\makeatletter
	\def\@captype{figure}
	\makeatother
	\includegraphics[scale=0.78]{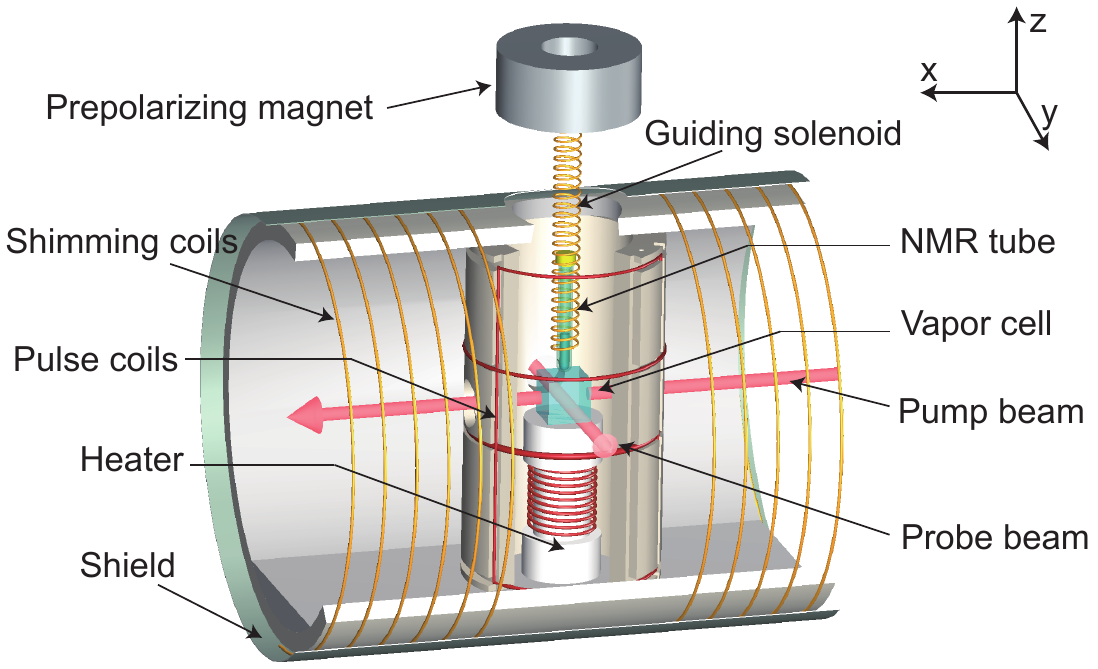}
	\caption{(color online). Schematic diagram of zero-field NMR spectrometer. The NMR sample, contained in a $5$-mm NMR tube, is detected with an atomic magnetometer with a $^{87}$Rb vapor cell operating at $155$~$^\circ$C.}
	\label{fig3}
\end{figure}

\begin{figure}[b]
	\makeatletter
	
	\def\@captype{figure}
	
	\makeatother
	\centering\includegraphics[scale=0.8]{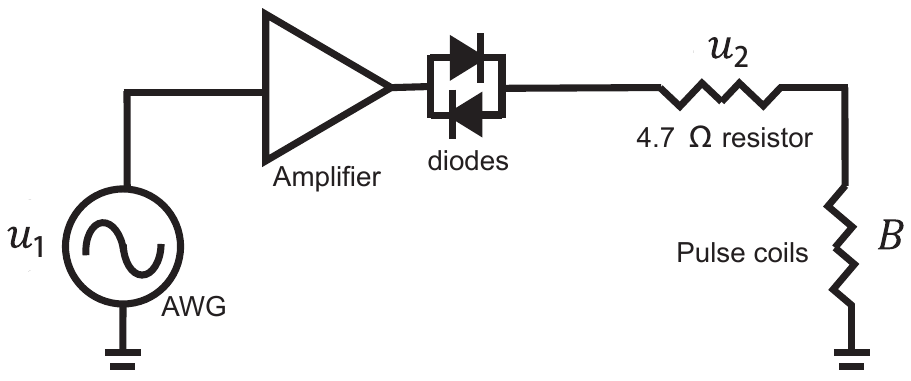}
	\caption{Pulse generation circuit. The anti-parallel diodes are placed in series with the pulse coils to isolate the noise of amplifier. The anti-parallel diodes introduce a voltage drop, which is measured in calibration experiments.}
	\label{circu}
\end{figure}

\subsection{Experimental details} 
TOC experiments were performed using a home-built zero-field NMR spectrometer, as illustrated in FIG. \ref{fig3}. Nuclear spins in the $^{13}$C-formic acid sample ($\approx 230~\mu L$) were polarized in a 1.3-T prepolarizing magnet,
after which the sample was shuttled into a magnetically shielded region,
such that the bottom of the sample tube is $\sim 1~\textrm{mm}$ 
above a $^{87}$Rb vapor cell of an atomic magnetometer \cite{Allred2002, Budker2007}. 
The $^{87}$Rb atoms in the vapor cell were pumped with a circularly polarized laser beam propagating in the $x$ direction.
The magnetic fields were measured via optical rotation of linearly polarized probe laser light at the $\textrm{D}2$ transition propagating in the $y$ direction.
The magnetometer was primarily sensitive to $z$ component of the nuclear magnetization, i.e., $M_z \propto  \textrm{Tr} [\rho(t) (\gamma_\textrm{H} \sigma_z \otimes \mathbf{1}_2+ \gamma_\textrm{C} \mathbf{1}_2 \otimes \sigma_z)]$, with a noise floor of about 15~$\textrm{fT}/\sqrt{\textrm{Hz}}$ above 100 Hz,
here $\rho(t)$ is the density matrix of the $^1$H-$^{13}$C system.A guiding magnetic field ($\approx 1~\textrm{G}$) was applied during the transfer,
and was adiabatically switched to zero after the sample reached  the zero-field region.
In our experiment, to ensure adiabaticity, the decay time to turn off the guiding field is 1 s.
Thus the spin system is initially  prepared in the adiabatic state \cite{Emondts2014}: $\rho (0) =\mathbf{1}_4/4+\frac{\epsilon_\textrm{H}+\epsilon_\textrm{C}}{2}(\sigma_z\otimes \mathbf{1}_2+ \mathbf{1}_2\otimes \sigma_z)-\frac{\epsilon_\textrm{H} - \epsilon_\textrm{C}}{4} (\sigma_x\otimes \sigma_x+\sigma_y \otimes \sigma_y)$ with the polarizations $\epsilon_\textrm{H}, \epsilon_\textrm{C} \sim 10^{-6}$.
The TOC pulses were generated by three sets of mutually orthogonal low-inductance pulse coils, which were individually controlled by arbitrary waveform generators (Keysight 33512B with two channels, Keysight 33511B with single channel),
and amplified individually with linear power amplifiers (AE TECHRON 7224) with 300 KHz bandwidth.

\begin{figure}[b]
	\makeatletter	
	\def\@captype{figure}	
	\makeatother
	\centering\includegraphics[scale=1]{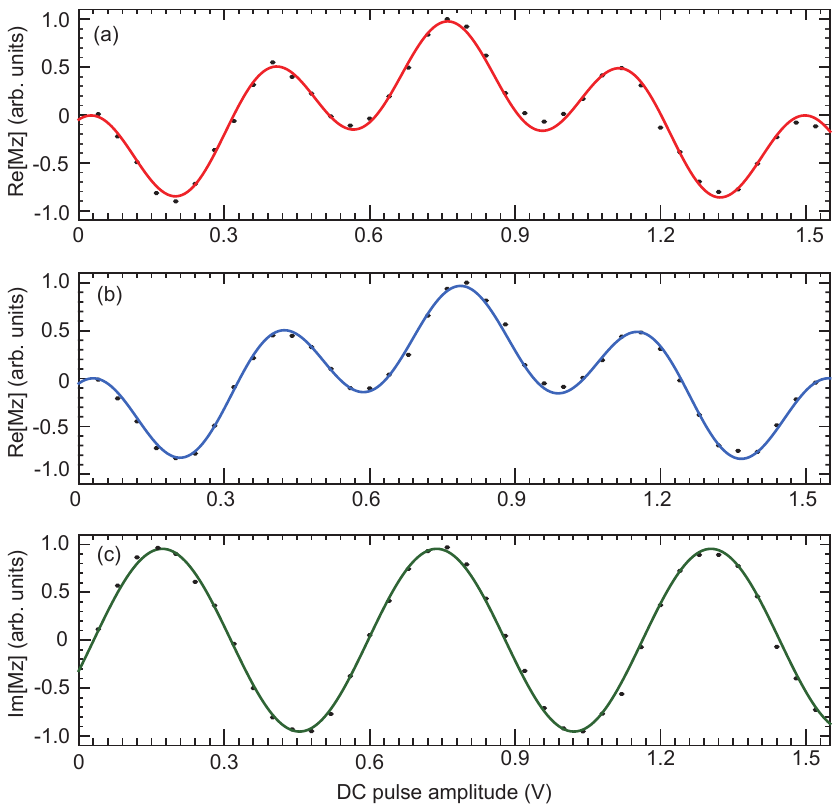}
	\caption{ (color online). Dependence of signal amplitude on the DC pulse amplitude for pulses of magnetic field in the $x$ (a), $y$ (b), and $z$ (c) directions. The solid curves overlaying the data are fits to curves described in the text.}
	\label{calib}
\end{figure}

\begin{figure*}
	\makeatletter
	
	\def\@captype{figure}
	
	\makeatother
	\centering\includegraphics[scale=0.8]{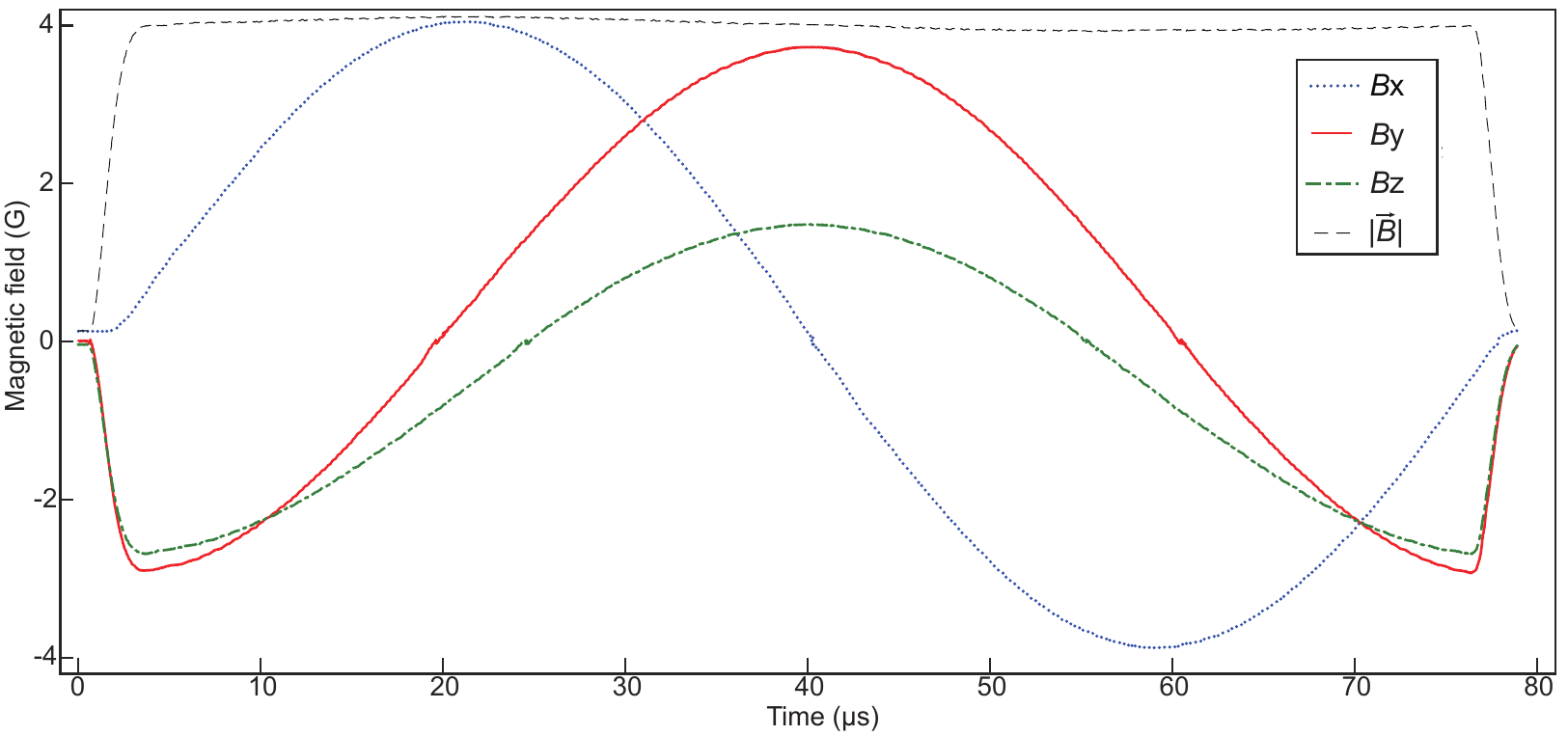}
	\caption{ (color online). Experimental result of measuring TOC $\pi$ pulse. TOC $\pi$ pulse is measured by acquiring the load voltage of the resistor in series with the pulse coils. }
	\label{shapes}
\end{figure*}

In FIG. \ref{circu} we present a scheme of the pulse generation circuit. FIG.  \ref{calib} describes how the signal  amplitude in various directions depends on the DC pulse amplitude. FIG. \ref{shapes} reports an example of the experimental optimal controls' shapes, in various directions.

\subsection{Performance of the TOC} 

\begin{figure}[b]  
	\makeatletter
	\def\@captype{figure}
	\makeatother
	\includegraphics[scale=1]{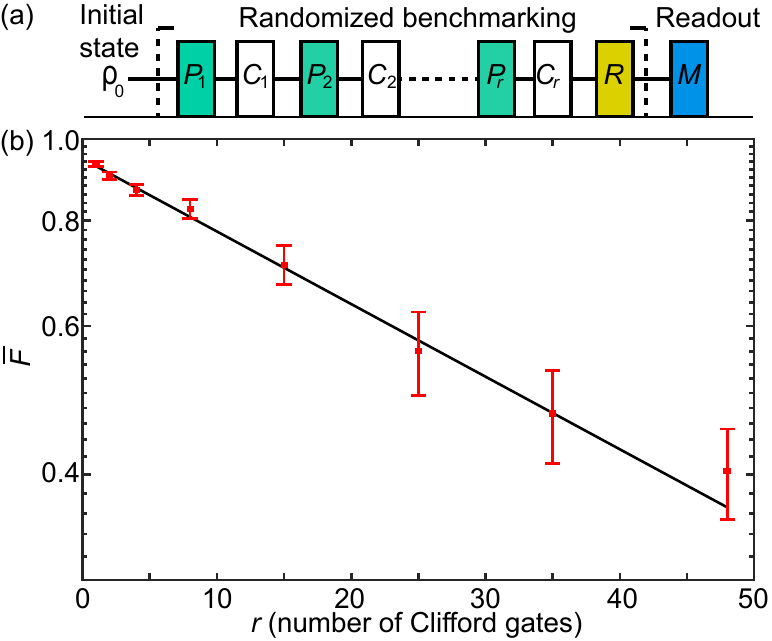}
	\caption{(color online). Randomized benchmarking (RB) experimental results.~(a) RB pulse sequences. (b) RB results for $^1$H single-spin TOC control. Each point $\bar{F}$ is an average over 32 random sequences of $r$ Clifford gates, and the error bars indicate the standard error of the mean. The $y$ axis is in log-scale. A single exponential decay 
	(solid line) fits the fidelity decay and reveals an average gate fidelity of $0.99$.}
	\label{fig4}
\end{figure}

To evaluate the quality of single-spin TOC control,
we adopted  the randomized benchmarking (RB) method \cite{Nielsen2002, Magesan2011}.
The RB pulse sequences are shown in Fig.~\ref{fig4}(a).
The initial state is prepared as $\rho_0 =\mathbf{1}_4/4+\frac{\epsilon_\textrm{H}}{2}\sigma_z\otimes \mathbf{1}_2+ \frac{\epsilon_\textrm{C}}{2} \mathbf{1}_2\otimes \sigma_z$
with the polarizations $\epsilon_\textrm{H}, \epsilon_\textrm{C} \sim 10^{-6}$.
Random sequences with $P=e^{\pm i \frac{\pi}{2} V} \otimes \mathbf{1}_2$ and $C=e^{\pm i \frac{\pi}{4}Q} \otimes \mathbf{1}_2$
are realized by TOC control, and
are applied for each sequence of length $m$,
where $V\in \{\mathbf{1}_2, \sigma_x, \sigma_y, \sigma_z \}$ and $Q \in \{\sigma_x, \sigma_y, \sigma_z \}$.
The Clifford gates are realized by combined operations $PC$.
The recovery gate $R$ is chosen to return the system to the initial state.
To measure the coefficient of $\sigma_z \otimes \mathbf{1}_2$ independently,
we adopted a recently developed state-tomography technique in zero-field NMR (see Ref.~\cite{Jiang2017}).
By averaging the coefficients of $\sigma_z \otimes \mathbf{1}_2$ over 32 different RB pulse sequences with the same length $r$,
and normalizing this averaged value to that of $r=0$,
the normalized signal $\bar{F}$ can be fitted by $\bar{F}=(1-d_{if})(1-2 \epsilon_{g})^r$,
where $d_{if}$ is due to the imperfection of the initial state preparation and readout,
and $\epsilon_{g}$ is the average gate error per Clifford gate.
As shown in Fig.~\ref{fig4}(b), the RB results yield an average gate error per Clifford gate $\epsilon_{g}=0.01$,
and an imperfection of the initial state preparation and readout $d_{if}=0.05$.
The average fidelity for $^1$H single-spin TOC control is $f_{avg}=1-\epsilon_{g}=0.99$.

 Errors in quantum control may be 
unitary, decoherent, and incoherent  \cite{Pravia2003}.
For our experiment, the most relevant is the unitary error  from pulse  distortion and miscalibration amplitude, caused by the bandwidth-limited pulse generation circuit, with
the pulse rise/fall time $\approx 5~\mu s$.
As the duration of TOC is shorter than that of composite pulse scheme, the rising edge will take a larger proportion in TOC pluses, hence cause more degradation in the control performance. This drawback due to the very short duration of TOC can be overcome through decreasing the total control amplitude (i.e., increasing the duration).
In the future, it may be possible to correct such pulse distortion  using a technique similar to the pre-distortion technique of \cite{Feng2016}.
The effect of $^1$H-$^{13}$C spin-spin interaction gives an error of $\approx 5\times 10^{-4}$ per gate.
The decoherent error, estimated to be $\sim 1\times 10^{-5}$ per gate, is even smaller since the coherence time of our system is substantially longer 
than the TOC pulse duration.
The incoherent error, which mainly comes from pulse-field inhomogeneity \cite{Jiang2017}, measured
to be $\sim 0.2\%$ over the sample volume, is estimated to be about $1\times 10^{-5}$ per gate.




For our $^1$H-$^{13}$C system with $ \frac{\gamma_\textrm{C}}{\gamma_\textrm{H}} \approx 0.2514=\gamma$, figure~\ref{fig01}(a) shows a $70-80 \% $ time gain of   TOC with respect to the composite-pulse scheme of \cite{Bian2017, Jiang2017}.  The reason for this gain is that the schemes of \cite{Bian2017, Jiang2017} do not consider time optimality. Moreover they use control fields in two directions only rather than three. 


\begin{figure}[b]  
	\makeatletter
	\def\@captype{figure}
	\makeatother
	\includegraphics[scale=1]{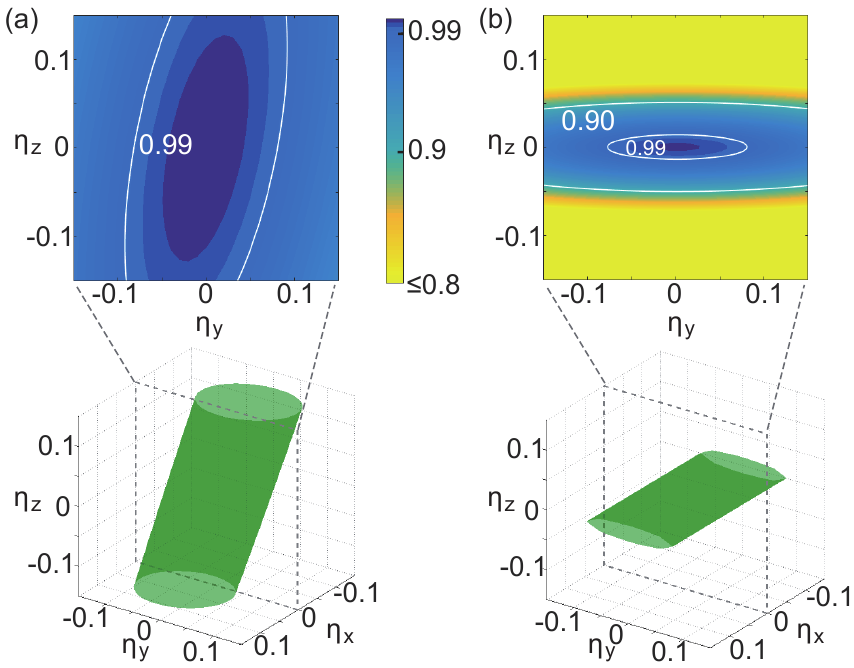}
	\caption{(color online). Robustness tests of TOC and composite pulse scheme with respect to pulse imperfections. The amplitude distortion ratio in $k=x,y,z$ is defined as $\eta_k=\frac{\textrm{amplitude distortion in k}}{\textrm{ideal value in k}}$. Lower figures in (a), (b) show the isosurface of $\textrm{fidelity}=0.99$ for TOC and composite pulse, respectively. Upper figures in (a), (b) are the fidelity contour maps for these two methods. 
		The better fidelity  of TOC is indicated by the larger area with a fidelity higher than $0.99$.}
	\label{iso}
\end{figure}

\subsection{Robustness of the optimal control law}\label{rob}

Once the optimal control problem is solved with a bound $L$ on the control and $t_{\textrm{min}}$, the number $Lt_{\textrm{min}}$ is independent of $L$ and it   gives the {\it sub-Riemannian distance} $d$ of the final condition $U_{f,1}\otimes U_{f,1}$ from the identity, $d:=d(U_{f,1} \otimes U_{f,2})$. According to Chow-Rashevskii theorem (see, e.g., \cite{ABB}) such a distance is equivalent to the given metric on the manifold (in this case $\textrm{SU}(2) \otimes \textrm{SU}(2)$). From a practical point of view it is interesting to investigate how robust the control law is with respect to the variations in the parameters of the model, in particular the parameter $\gamma$. Assume for instance that the gyromagnetic ratio $\gamma_1$
is known with some confidence while $\gamma_2$, and therefore $\gamma$ is known with less confidence.
From Theorem 1 it follows that the final value on the first spin is independent of $\gamma$ while the error is all on the final state of the second spin.
Differentiating the operation $U_2$ of Theorem 1, i.e., the operation on the second spin, with respect to $\gamma$ we obtain (for $U_2=U_{f,2}$, the desired final condition)
\be{differnt678}
\frac{d U_{f,2}}{d\gamma}=e^{A t_{\textrm{min}}} P t_{\textrm{min}} e^{-A t_{\textrm{min}}} U_{f,2},
\ee
so that (assuming because of unitarity $\|U_{f,2}\|=1$), $\|\frac{d U_{f,2}}{d\gamma}\|\leq \|e^{A t_{\textrm{min}}} Pt_{\textrm{min}} e^{-At_{\textrm{min}}}\|=|P|t_{\textrm{min}}:=Lt_{\textrm{min}}=d(U_{f,1} \otimes U_{f,2})$.  Therefore, the sensitivity with respect to $\gamma$ of the final condition is bounded by the sub-Riemannian distance
of the desired  final condition.
From simulation, a $1 \%$ deviation in $\gamma$ will only result in a $0.001\%$ drop in fidelity (for the TOC used in this experiment). 

The robustness of TOC against distortions in control fields is also demonstrated in \figref{iso}. Even in this regard, the TOC is preferable as compared to composite pulse scheme.

\section{Conclusion and discussion} \label{sec_6}

We have theoretically derived  and  experimentally 
demonstrated, the time optimal control of two independent spin-$\frac{1}{2}$ 
particles by simultaneous control. Novel techniques of symmetry reduction allowed us to obtain {analytic} expressions of the TOC,   {with minimal use of  numerical experiments}.  Such  control fields, implemented using a zero-field heteronuclear NMR system, gave an average
fidelity of $99\%$, and considerable time saving. Our paper adds to the recently growing literature that combines {\it analytical} methods with {\it experimental} implementations in quantum mechanical control \cite{Geng2016}, \cite{KhanP1}, \cite{KhanP2}, \cite{KhanP3}, 
\cite{Avinadav2014}, \cite{glaserpraref}.

Typical optimal control techniques and applications to quantum systems use numerical methods which involve the repeated {\it numerical integration} of a system of differential equations with variable initial conditions (parameters). In our case, there is no need of numerical integration since the solution is given in explicit form. Moreover the number of parameters is reduced to a minimum with the technique of symmetry reduction. Still computer experiments  can be a useful tool to solve the Task 3 of the procedure in section \ref{Proce} by visualizing the reachable set in the quotient space and-or by helping in the solution of integer optimization problems, such as the one described in Theorem \ref{OptimizationProblem}.  

Ideas presented here can be applied more in general for quantum systems displaying symmetries such as the $KP$ systems considered in \cite{Albertini2018}. The analytic knowledge of the TOC is useful even in cases where such a control is not the one physically implemented. It gives information about the inherent time limitations of the system, therefore indicating a benchmark for the time of  any control law. The knowledge of the TOC law for any final condition is also equivalent to a description of the reachable sets which, in the presence of symmetries, can be carried out in the (reduced) orbit space \cite{Albertini2018}. 
 
It is interesting to investigate whether the optimal control techniques discussed here can be scaled to higher dimensional systems and in particular in the simultaneous control of $N>2$ spin $\frac{1}{2}$ systems. In general, optimal control problems become harder as the dimension of the system increases with respect to the number of controls. More specifically, the main reason why we were able to find an explicit form for the optimal control and trajectory for our system is the fact that the system has {\it degree of non-holonomy} one, that is, it is enough to do {\it one}  Lie bracket of the vector fields which appear in the Schr\"odinger equation to obtain the whole 
Lie algebra of available directions of motion. This property is lost if  we increase the number of spins and keep the number of control fixed. It maybe recovered if we introduce additional controls by, for example, assuming  that $N/2$ systems each consisting of two spin $\frac{1}{2}$'s can be controlled independently. Under these assumptions, one may use techniques similar to the ones considered in \cite{Domenico1}, \cite{Domenico2} for the case of $N/2$ systems each consisting of one spin only.  Such controls and optimal times still give lower bounds on the time of transfer in more realistic scenarios with fewer controls.


\textsl{Acknowledgment.\textbf{--}}We thank X. Rong and A. Voronov for valuable discussions. Support came from National Key Research and Development Program of China (Grant No. 2018YFA0306600), National Key Basic Research Program of China (Grant No. 2014CB848700), National Natural Science Foundation of China (Grants Nos. 11425523, 11375167, 11661161018, 11227901), Anhui Initiative in Quantum Information Technologies (Grant No. AHY050000), National Science Foundation (Grant ECCS 1710558). 

Y. J., J. B., and M. J. contributed equally to this work.

\section{appendix} \label{APP1}

\subsection{Proof of Proposition \ref{SU2orbits}}\label{proofchar}
{\it Proof.}  We first have to show that the map (\ref{Psimap}) is well defined, i.e., it
does not depend on the choice of the representative $(U,Z)$ in $[(U,Z)]$ nor on the choice of $S$ when writing $U$ as $U=S \Lambda S^\dagger$ in (\ref{Psi1}). Let us show the latter property first. If $U=S_1\Lambda S_1^\dagger=S_2\Lambda S_2^\dagger$ then $S_1^\dagger S_2$ commutes with $\Lambda$ and since $\Lambda$ is not $\pm {\bf 1}$ then $S_1^\dagger S_2$ must be a diagonal matrix. This implies that we must have
$(S_1^\dagger Z  S_1)_{1,1}=(S_2^\dagger Z  S_2)_{1,1}$. Consider now (\ref{Psi1}). Assume that instead of $(U,Z)$ we take the representative $(FUF^\dagger,FZF^\dagger)$, in $[(U,Z)]$ for some $F \in SU(2)$. Write $FUF^\dagger :=  FS \Lambda S^\dagger F^\dagger$. This gives the same value of $\phi$ and it does not change the $(1,1)$ entry of $(S^\dagger F^\dagger) F Z F^\dagger (FS)=S^\dagger Z S$. Analogously in the case (\ref{Psi2}) and (\ref{Psi3}) a similarity transformation does not modify the eigenvalues of $X$ and $Z$.

It is clear that the map $\Psi$ is onto. To show that it is one to one assume that
$\Psi([(U_1, Z_1)])=\Psi([(U_2, Z_2)])$. Then the eigenvalues of $U_1$ and $U_2$ are the same. If $U_1$ and $U_2$ are both $\pm {\mathbbm{1}}$ then the eigenvalues of $Z_1$ and $Z_2$ must be the same,  and so there exists an $F \in \textrm{SU}(2)$ such that
$Z_2=FZ_1 F^\dagger$. If $U_1$, and therefore $U_2$, is
different from $\pm {\mathbbm{1}}$, they have however the same eigenvalues. So there exist $S_1, S_2  \in \textrm{SU}(2)$ such that $U_1 =S_1\Lambda S_1^\dagger$ $U_2=S_2\Lambda S_2^\dagger$. Moreover $S_1^\dagger Z_1 S_1$ and  $S_2^\dagger Z_2 S_2$ have the same $(1,1)$ entry. So they only differ by similarity transformation by a diagonal matrix $H_d$  and we have $S_2^\dagger Z_2 S_2=H_d S_1^\dagger Z_1 S_1 {H_d}^\dagger$, from which
\be{equi1}
Z_2=S_2 H_dS_1^\dagger Z_1 S_1 {H_d}^\dagger S_2^\dagger.
\ee
By writing $U_1=S_1 \Lambda S_1^\dagger$ as $U_1=S_1 {H_d}^\dagger  \Lambda {H_d} S_1^\dagger$ and from $U_2=S_2 \Lambda S_2^\dagger$,  we obtain
\be{equi2}
U_2=S_2 {H_d} S_1^\dagger U_1 S_1 {H_d}^\dagger S_2^\dagger.
\ee
Comparing (\ref{equi1}) and (\ref{equi2}), we have that $[(U_1,Z_1)]=[(U_2,Z_2)]$.

\subsection{Some useful computations}\label{UsC}

We report here the results of some computations, in particular the exponential of matrices,  which are useful in the process of determining the optimal control.
We first calculate the exponential $e^{Ft}$ with
\be{espoS}
F:=ic\sigma_z-id\sigma_y=
\begin{pmatrix}i c & -d \cr
	d & -ic  \end{pmatrix} .
\ee
We have
\begin{widetext}
		\be{espo3}
e^{Ft}=\begin{pmatrix} \cos(\sqrt{c^2+d^2} t)+i \frac{c}{\sqrt{c^2+d^2}}\sin(\sqrt{c^2+d^2}t) & -\frac{d}{\sqrt{c^2+d^2}}\sin( \sqrt{c^2+d^2}t) \cr
\frac{d}{\sqrt{c^2+d^2}}\sin( \sqrt{c^2+d^2}t) & \cos(\sqrt{c^2+d^2} t)-i \frac{c}{\sqrt{c^2+d^2}}\sin(\sqrt{c^2+d^2}t)\end{pmatrix}.
\ee	
\end{widetext}

This formula can be used to compute $\tilde{U}_1(t)$ and $\tilde{U}_2(t)$ in (\ref{U12}) 
by using, for the second factor of
\be{X1uuu}
\tilde{U}_1=e^{i\omega \sigma_z t} e^{(ia \sigma_z - ib \sigma_y-i\omega \sigma_z)t}=
e^{i\omega \sigma_z t} e^{[i(a-\omega)\sigma_z-ib\sigma_y]t},
\ee
\be{toBus2}
c:=a-\omega, \qquad d:=b,
\ee
and, for the second factor of
\begin{equation}\label{X1uuur}
\begin{aligned}
\tilde{U}_2&=e^{i\omega \sigma_z t} e^{(i\gamma a \sigma_z -i \gamma b \sigma_y-i\omega \sigma_z)t}\\
&=e^{i\omega \sigma_z t} e^{[i(\gamma a-\omega)\sigma_z-i\gamma b\sigma_y]t},
\end{aligned}
\end{equation}
\be{toBus3}
c:=\gamma a-\omega, \qquad d:=\gamma b.
\ee
Let us consider the case of $\tilde{U}_2$ since the case of $\tilde{U}_1$ can be recovered by setting $\gamma=1$.
In this case, a simple calculation using the fact that $a^2+b^2=1$ and formulas (\ref{toBus3}) gives

\begin{equation}\label{uu3}
\begin{aligned}
\sqrt{c^2+d^2}&=\sqrt{(\gamma a-\omega)^2+\gamma^2\left(1-a^2\right)}\\
&=\sqrt{\omega^2+\gamma^2-2a\omega \gamma}:=\eta_\gamma:=\eta_\gamma(a,\omega).
\end{aligned}
\end{equation}

Let us assume $|a|<1$ so that $\eta_1 \not=0$ and, in general, $\eta_\gamma \not=0$ for any $\gamma$. Then using (\ref{espo3}), we have for $\tilde{U}_1$ in (\ref{X1uuu})
\begin{widetext}
\be{X1iii}
\tilde{U}_1(t):=\begin{pmatrix} e^{i\omega t} (\cos(\eta_1 t)+i\frac{a-\omega }{\eta_1} \sin(\eta_1 t)) & -e^{i\omega t}\frac{b}{\eta_1} \sin(\eta_1 t) \cr
	e^{-i\omega t}\frac{b}{\eta_1} \sin(\eta_1 t) & e^{-i\omega t} (\cos(\eta_1 t)-i\frac{a-\omega}{\eta_1} \sin(\eta_1 t))
\end{pmatrix}.
\ee
\end{widetext}
Also, still when $|a|<1$, using (\ref{espo3}), we have for $\tilde{U}_2$ in (\ref{X1uuur})
\begin{widetext}
\be{X2iii2}
\tilde{U}_2(t):=\begin{pmatrix} e^{i\omega t} (\cos(\eta_\gamma t)+i\frac{\gamma a-\omega }{\eta_\gamma} \sin(\eta_\gamma  t)) & -e^{i\omega t}\frac{\gamma b}{\eta_\gamma} \sin(\eta_\gamma t) \cr
	e^{-i\omega t}\frac{\gamma b}{\eta_\gamma} \sin(\eta_\gamma t) & e^{-i\omega t} (\cos(\eta_\gamma t)-i\frac{\gamma a-\omega }{\eta_\gamma} \sin(\eta_\gamma  t)),
\end{pmatrix}
\ee
\end{widetext}

\subsection{Solution of the Optimization Problem of Theorem \ref{OptimizationProblem} for the case $\gamma=1/2$}\label{gamma12}
Consider $\gamma=\frac{1}{2}$ and $\theta=\pi$ ($q=1$). The desired evolution is
\be{DefiNOT1}
U_{f,1}=\pm\textrm{Phase}=\pm \begin{pmatrix} -i & 0 \cr 0 & i \end{pmatrix},  \quad U_{f,1}=\pm \begin{pmatrix} 1 & 0 \cr 0 & 1 \end{pmatrix}
\ee
With $\theta=\pi$, \eqref{Req3} can be simplified to
\begin{equation}
\label{newml}
m\sqrt{\omega^2+1 -2a\omega} \,t= \frac{{l} \pi}{2},\quad
\end{equation}
where $l$ can only be an odd positive integer.
Thus the cases with $s=\pm 1$ can be combined, and \eqref{emmegamma} becomes
\begin{equation}
\label{newm}
M(m,l,k):=m^2(1-\gamma)+\frac{l^2}{4} \gamma-k^2,
\end{equation}
with the constraint \eqref{Req10} simplified to
\begin{equation}
\label{newcon}
\left(m-\frac{l}{2}\right)^2 < \frac{M(m,l,k)}{\gamma (1-\gamma)} < \left(m+\frac{l}{2}\right)^2.
\end{equation}
\noindent Using \eqref{newm} and \eqref{newcon}, we obtain the condition
\be{Req11}
\left( m- \frac{l}{2}\right)^2 < 4k^2 < \left( m+ \frac{l}{2}\right)^2.
\ee
Given $M_\gamma$ in \eqref{newm}, (\ref{time}), and the fact that
$\gamma(1-\gamma)$ is positive in this case,  we need to find the largest possible $k$ which satisfies (\ref{Req11}) in terms of $l$ and $m$. If we define $\tilde{s}:=2m+l$, the largest possible $k$ which satisfies the right  inequality  in (\ref{Req11}) is (recall $l$ is odd so $\tilde{s}=2m+l$ is also odd)
\be{Seven}
k=\frac{\tilde{s}-1}{4}, \qquad \texttt{for} \qquad \tilde{s}=5,9,13,...,
\ee
and
\be{Sodd}
k=\frac{\tilde{s}-3}{4}, \qquad \texttt{for} \qquad \tilde{s}=7,11,15,...
\ee
The case $\tilde{s}=3$ is not possible since $k\not=0$. Also with this choice of $k$ the left inequality in (\ref{Req11}) is always satisfied unless $l=1$ and $m$ is odd. In the latter case, since $k$ in (\ref{Sodd}) and (\ref{Seven}) is the largest possible value, no other value of $k$ would satisfy the left inequality in (\ref{Req11}). Thus, the values $l=1$ and $m$ odd are excluded from the search. Replacing $k$ in (\ref{newm}), with $\gamma=\frac{1}{2}$,  we obtain
\be{NewM}
\begin{aligned}
	\frac{M}{\gamma(1-\gamma)}&=2m^2+\frac{l^2}{2}-4k^2=\frac{1}{2}(2m+l)^2-4k^2-2ml\\
	&=\frac{\tilde{s}^2}{2}-\frac{(\tilde{s}-h)^2}{4}-2m(\tilde{s}-2m)\\
	&=\frac{\tilde{s}^2}{4}-\frac{(\tilde{s}-h)^2}{4}+\left(2m-\frac{\tilde{s}}{2} \right)^2,
\end{aligned}
\ee
where $h=1$ or $h=3$ if we are in case (\ref{Seven}) and (\ref{Sodd}) respectively.
For a given $s$ the minimum in (\ref{NewM}) is given for $m=\frac{\tilde{s}-h}{4}$, and it is given by
\be{NewMopt}
\frac{M}{\gamma(1-\gamma)}=\frac{h\tilde{s}}{2},
\ee
which is minimized with $\tilde{s}=5$ ($h=1$). We have therefore the following optimal
values for $m$, $l$ and $k$ (with $\tilde{s}=5$ and $h=1$):
\be{valuesmlk}
m=\frac{\tilde{s}-h}{4}=1, \ l=\tilde{s}-2m=3, \ k =  \frac{\tilde{s}-h}{4}=1.
\ee


\begin{thebibliography}{99}
\bibitem{footnote2}
It is a general fact that in driftless TOC 
	problem such as the one considered here the bound $|\vec u|\leq D$, a.e., is equivalent 
	to $|\vec u|=D$. A formal proof of this fact is given in \cite{Albertini2018}.
\bibitem{footnote3}
This is the space of equivalence classes in $SU(2) \otimes SU(2)$ where $Z_1$ is equivalent to $Z_2$ if and only if there exists a 
	$\tilde Y \in G$ such that $\tilde YZ_1 \tilde Y^\dagger=Z_2$.
\bibitem{voronov} This result was proved in collaboration with A. Voronov at the School of Mathematics, University of Minnesota
\end{thebibliography}
\end{document}